\documentclass[12pt,eqno]{article}
\usepackage{amsmath,amssymb}
\numberwithin{equation}{section}

\newcommand{\alp}{\alpha}
\newcommand{\bt}{\beta}
\newcommand{\gm}{\gamma}

\newcommand{\dlt}{\delta}

\newcommand{\tht}{\theta}
\newcommand{\btht}{\bar{\tht}}
\newcommand{\vtht}{\vartheta}
\newcommand{\kp}{\kappa}
\newcommand{\lmd}{\lambda}

\newcommand{\sgm}{\sigma}
\newcommand{\Sgm}{\Sigma}
\newcommand{\vph}{\varphi}

\newcommand{\Omg}{\Omega}

\newcommand{\ztR}{\zeta_{\rm R}}

\newcommand{\be}{\begin{equation}}
\newcommand{\ee}{\end{equation}}
\newcommand{\bea}{\begin{eqnarray}}
\newcommand{\eea}{\end{eqnarray}}
\newcommand{\eql}{\!\!\!&=\!\!\!&}

\newcommand{\defa}{\!\!\!&\equiv\!\!\!&}

\newcommand{\mtrx}[4]{\brkt{\begin{array}{cc}#1&#2\\#3&#4\end{array}}}
\newcommand{\dgnl}[2]{\brkt{\begin{array}{cc}#1& \\ &#2\end{array}}}

\newcommand{\simgt}{\stackrel{>}{{}_\sim}}

\newcommand{\tl}[1]{\tilde{#1}}
\newcommand{\bdm}[1]{{\mbox{\boldmath $#1$}}}

\newcommand{\diag}{{\rm diag}}
\newcommand{\der}{\partial}
\newcommand{\dr}{\!\!d}
\newcommand{\hc}{{\rm h.c.}}
\newcommand{\ie}{{\it i.e.}}

\newcommand{\vev}[1]{\langle #1 \rangle}

\newcommand{\brkt}[1]{\left( #1 \right)}
\newcommand{\brc}[1]{\left\{ #1 \right\}}
\newcommand{\sbk}[1]{\left[ #1 \right]}
\newcommand{\abs}[1]{\left| #1 \right|}

\renewcommand{\Re}{{\rm Re}}
\renewcommand{\Im}{{\rm Im}}

\newcommand{\cA}{{\cal A}}
\newcommand{\cD}{{\cal D}}
\newcommand{\cF}{{\cal F}}
\newcommand{\cH}{{\cal H}}
\newcommand{\cI}{{\cal I}}
\newcommand{\cK}{{\cal K}}
\newcommand{\cL}{{\cal L}}

\newcommand{\cN}{{\cal N}}
\newcommand{\cO}{{\cal O}}

\newcommand{\cR}{{\cal R}}

\newcommand{\cV}{{\cal V}}
\newcommand{\cW}{{\cal W}}

\renewcommand{\ge}[2]{e_{#1}^{\;\;#2}}
\newcommand{\udl}[1]{\underline{#1}}
\newcommand{\nV}{n_{\rm V}}
\newcommand{\nH}{n_{\rm H}}
\newcommand{\dmx}{d_{\alp}^{\;\;\bt}}

\newcommand{\gey}{\vev{\ge{y}{4}}}
\newcommand{\SUu}{SU(2)_{\mbox{\scriptsize $\bdm{U}$}}}
\newcommand{\Usp}{USp(2,2\nH)}
\newcommand{\lrder}{\stackrel{\leftrightarrow}{\partial}}
\newcommand{\tlg}{\mbox{\scriptsize $\bdm{\tl{g}}$}}
\newcommand{\vpcl}{\varphi_{\rm cl}}
\newcommand{\vpR}{\varphi_{\rm R}}
\newcommand{\vpI}{\varphi_{\rm I}}

\newcommand{\NP}[1]{{\it Nucl.~Phys.}~{\bf #1}}
\newcommand{\PL}[1]{{\it Phys.~Lett.}~{\bf #1}}

\newcommand{\PR}[1]{{\it Phys.~Rev.}~{\bf #1}}
\newcommand{\PRL}[1]{{\it Phys.~Rev.~Lett.}~{\bf #1}}
\newcommand{\PTP}[1]{{\it Prog.~Theor.~Phys.}~{\bf #1}}

\newcommand{\JH}[1]{{\it JHEP}~{\bf #1}}

\begin{document}

\begin{titlepage}
\null
\begin{flushright}
 {\tt hep-th/0501183}\\
KAIST-TH 2004/22
\\
January, 2005
\end{flushright}

\vskip 2cm
\begin{center}
\baselineskip 0.8cm
{\LARGE \bf Dynamical radion superfield in 5D action}

\lineskip .75em
\vskip 2.5cm

\normalsize

{\large\bf Hiroyuki Abe}{\def\thefootnote{\fnsymbol{footnote}}
\footnote[1]{\it e-mail address:abe@muon.kaist.ac.kr}}
{\large\bf and Yutaka Sakamura}{\def\thefootnote{\fnsymbol{footnote}}
\footnote[2]{\it e-mail address:sakamura@muon.kaist.ac.kr}}

\vskip 1.5em

{\it Department of Physics, \\
Korea Advanced Institute of Science and Technology \\
Daejeon 305-701, Korea}

\vspace{18mm}

{\bf Abstract}\\[5mm]
{\parbox{13cm}{\hspace{5mm} \small
We derive 5D $\cN=1$ superspace action including the radion superfield. 
The radion is treated as a dynamical field and identified as 
a solution of the equation of motion even in the presence of 
the radius stabilization mechanism. 
Our derivation is systematic and based on the superconformal 
formulation of 5D supergravity. 
We can read off the couplings of the dynamical radion superfield 
to the matter superfields from our result. 
The correct radion mass can be obtained by calculating the radion potential 
from our superspace action. 
}}

\end{center}

\end{titlepage}

\clearpage

\section{Introduction}
Five dimensional supergravity (5D SUGRA) compactified 
on an orbifold~$S^1/Z_2$ has been thoroughly investigated 
since it is shown to appear as an effective theory of 
the strongly-coupled heterotic string theory~\cite{HW} compactified 
on a Calabi-Yau 3-fold~\cite{LOSW}. 
Especially, the Randall-Sundrum model~\cite{RS} is attractive 
as an alternative solution to the hierarchy problem, 
and a huge number of researches on this model have been done.  
In this model, the background geometry is a slice of the anti-de Sitter 
(AdS) spacetime and the metric has the form of \footnote{
Throughout this paper, we will use $\mu,\nu,\cdots=0,1,2,3,4$ for 
the 5D world vector indices, and $m,n,\cdots=0,1,2,3$ for the 4D indices. 
The coordinate of the extra dimension is denoted as $y\equiv x^4$. }
\bea
 ds^2 \eql g_{\mu\nu}dx^\mu dx^\nu 
 = e^{-2ky}\eta_{mn}dx^m dx^n-dy^2 \nonumber\\ 
 \eql e^{-2kR\vtht}\eta_{mn}dx^m dx^n-R^2 d\vtht^2, 
\eea
where $k$ is the AdS curvature and $R$ is the radius of the orbifold. 
The physical range of the extra space is $0\leq y \leq \pi R$. 
In the second line, we have changed the coordinate~$y$ 
to the dimensionless coordinate $\vtht\equiv y/R$. 

In such a brane-world model, the radius of the compactified extra dimension 
is generically a dynamical degree of freedom, {\it the radion}. 
In the original Randall-Sundrum model~\cite{RS}, the radius of the orbifold 
is undetermined by the dynamics and thus the radion is a massless field. 
Hence, it remains to be a dynamical degree of freedom 
in low energies and should be taken into account in 4D effective theory. 
A naive way of introducing the radion mode into the theory is 
to promote the radius~$R$ to a 4D field~$r(x)$, 
that is, to consider the metric of the form~\cite{GW2,CGRT} 
\be
 ds^2 = e^{-2kr(x)\vtht}g^{(4)}_{mn}(x)dx^m dx^n-r^2(x)d\vtht^2, 
 \label{naive_radion}
\ee
where $g^{(4)}_{mn}$ is the 4D graviton. 
However, this is not a solution of the Einstein equation, 
even at the linearized order. 
This means that the radion mode defined here 
is not a mass eigenstate and has mixings with the massive Kaluza-Klein (K.K.) 
modes which we have dropped in Eq.(\ref{naive_radion}). 
Thus, such K.K. modes cannot simply be dropped when they are integrated out. 
Therefore, a naive ansatz~(\ref{naive_radion}) need to be corrected. 
Furthermore, this metric means the radion~$r(x)$ does not couple 
to the brane at $\vtht=0$, which contradicts the fact 
that it couples to the boundary branes 
like a Brans-Dicke scalar~\cite{CGRT,GT}. 
In order to treat the radion as a dynamical field, we have to define it 
as a solution of the equation of motion. 
Then, we can safely drop the K.K. modes after we integrate them out.
Such treatment for the radion is discussed in Ref.~\cite{CGR} 
in the absence of the radius stabilization mechanism. 
In that case, the physical gravitational modes are the 4D graviton, 
its K.K. modes, and the radion mode. 
Note that there is no K.K. tower above the massless radion. 
All such K.K. modes can be gauged away and thus are unphysical. 
%
The authors of Ref.~\cite{CGR} found a gauge where the radion mode is 
contained in the metric as 
\be
 ds^2 = e^{-2ky+\tl{b}(x)e^{2ky}}g_{mn}^{(4)}(x)dx^m dx^n
 -\brkt{1-\tl{b}(x)e^{2ky}}^2 dy^2,  \label{CGR_metric}
\ee
where $\tl{b}(x)$ is a 4D massless field that corresponds to 
the radion fluctuation mode around the background value. 
In contrast to the naive ansatz~(\ref{naive_radion}), 
this certainly satisfies the linearized Einstein equation. 
(Note that the mode function of $\tl{b}(x)$ is the correct 
one, $e^{2ky}$.) 
This gauge is useful because the whole spacetime is covered by 
one coordinate patch and simultaneously the 4D boundary planes are 
expressed by constant values of $y$.\footnote{
For example, the Gaussian normal coordinates, which is convenient 
to deal with the junction conditions at the boundaries, 
need two coordinate patches to cover the whole spacetime, 
and in the Newton gauge, which is convenient to discuss the bulk dynamics, 
the boundary planes are no longer expressed by constant values of $y$. 
}

In order to construct a realistic model, we have to introduce some 
stabilization mechanism for the radius~$R$. 
One of the main stabilization mechanism is proposed in Ref.~\cite{GW}, 
and it involves a bulk scalar field 
that has a nontrivial vacuum configuration. 
In such a case, the metric receives the backreaction from the bulk scalar 
configuration, and the background geometry deviates from the AdS spacetime. 
The radion dynamics in this case is thoroughly investigated 
in Ref.~\cite{CGK}. 
In this case, the radion mode resides not only in the metric 
but also in the bulk scalar field that is relevant to
the radius stabilization. 

The supersymmetric extension of the Randall-Sundrum model 
has also been investigated in many papers~\cite{SUSY_RS}. 
In this case, the radion belongs to an $\cN=1$ chiral multiplet 
in 4D effective theory. 
The corresponding radion superfield is identified in Ref.~\cite{BNZ,LLP} 
in the absence of the bulk matter fields. 
The couplings between the radion superfield and the bulk matter superfields 
are provided in Ref.~\cite{MP}, 
but their derivation is based on the naive ansatz~(\ref{naive_radion}) 
and should be corrected. 
There are also works that try to identify the radion multiplet 
in the context of the superconformal gravity~\cite{KO1,CST}. 
In Ref.~\cite{KO1}, a chiral multiplet that contains the extra component 
of the f\"{u}nfbein~$\ge{y}{4}$ is constructed from component fields of 
5D superconformal multiplets. 
The authors of Ref.~\cite{CST} have clarified the appearance of 
such a chiral multiplet, which is usually called the radion multiplet, 
in the $\cN=1$ description of the superconformal gravity action. 
However, this multiplet is not the radion multiplet itself 
although it is closely related to the latter, 
because $\ge{y}{4}$ is a 5D field while the radion is a 4D field. 
In order to clarify the relation between them, we have to solve 
the equations of motion. 
The purpose of this paper is to derive 5D action written by 
$\cN=1$ superfields including the dynamical radion superfield 
defined as a solution of the equations of motion, 
in an appropriate way. 
Roughly speaking, our work corresponds to an extension of Ref.~\cite{BNZ} 
to the case where the bulk and boundary matters 
and the radius stabilization mechanism exist. 

In our previous paper~\cite{AS}, we have derived 5D superspace action 
on a general warped background directly from 5D SUGRA action.\footnote{
The same superspace action is also obtained in Ref.~\cite{CST} 
independently of our work, 
at the stage before the superconformal gauge fixing. } 
In this work, however, we have fixed the gravitational multiplet 
to its background value, and dropped all the fluctuation modes 
including the radion. 
Thus, we will derive the desired 5D action by 
introducing the dynamical radion mode in the 5D superspace action 
obtained in our previous work. 

The paper is organized as follows. 
In the next section, we will briefly review the discussion in Ref.~\cite{CGK} 
to understand the situation, 
and explicitly identify the radion mode 
as a solution of the linearized equation of motion 
in the model of Ref.~\cite{MO} as an example. 
In Section~\ref{rad_sf_action}, we will derive the desired superspace action, 
and clarify the couplings of the radion superfield to the matter superfields. 
Section~\ref{summary} is devoted to the summary. 
We collect the equations held by the classical background solution 
in Appendix~\ref{EOM_bg}, and give some comment on the Newton gauge 
in Appendix~\ref{bd_cond_Ng}. 
In Appendix~\ref{sf_gf}, we collect explicit forms of the superfields 
in terms of the superconformal notation of Ref.~\cite{KO1,KO2}.

\section{Dynamical radion mode}
In this section, we will identify the dynamical radion mode 
in the presence of the stabilization mechanism. 
Some results of this section are contained in Refs.~\cite{CGK,MO,EMS}. 
We will provide a self-contained review before deriving 
the desired superspace action 
since it is helpful to understand the situation around the radion. 

\subsection{Linearized equations of motion}
First, we will derive the linearized equations of motion 
in order to discuss the dynamical radion mode. 

For simplicity, we will assume that only one bulk scalar field has 
a nontrivial vacuum configuration in the following. 
Then, the lagrangian is \footnote{ 
The metric convention is $\eta_{\mu\nu}=\diag(1,-1,-1,-1,-1)$. 
} 
\be
 \cL = -M_5^3\sqrt{g}\cR + \sqrt{g}\brc{
 \frac{1}{2}\der^\mu\bar{\vph}\der_\mu\vph-V(\vph)}
 +\sqrt{\abs{g_0}}\lmd_0(\vph)\dlt(y)
 +\sqrt{\abs{g_\pi}}\lmd_\pi(\vph)\dlt(y-\pi R)
 +\cdots, 
\ee
where $M_5$ is the 5D Planck mass, $g\equiv \det(g_{\mu\nu})$ 
and $\cR$ is the 5D Ricci scalar. 
$V(\vph)$ and $\lmd_0(\vph)$, $\lmd_\pi(\vph)$ are the scalar potentials 
in the bulk and on the boundaries respectively, and 
$g_0$ and $g_\pi$ are the determinants of the induced metrics 
on the branes. 
The ellipsis denotes terms irrelevant to the radius stabilization. 

The background that preserves 4D Poincar\'{e} invariance is 
\bea
 ds^2 \eql e^{2\sgm(y)}\eta_{mn}dx^m dx^n-dy^2, \nonumber\\
 \vph \eql \vpcl(y). \label{bkgd}
\eea
The equations of motion and the jump conditions for this background 
are listed in Appendix~\ref{EOM_bg}. 

In order to discuss the dynamical degrees of freedom, 
we should solve the linearized 5D Einstein equation 
and the field equations. 
For this purpose, the Newton gauge is useful \cite{EMS,montes}. 
In this gauge, the fluctuation modes around the background are parametrized as 
\bea
 ds^2 \eql e^{2\sgm}\brkt{\eta_{mn}+h^{\rm TT}_{mn}+2B\eta_{mn}}dx^m dx^n
 -\brkt{1-4B}dy^2, \nonumber\\
 \vph \eql \vpcl+\tl{\vph},  \label{NT_gauge}
\eea
where $h^{\rm TT}_{mn}$ is the transverse traceless mode, \ie, 
$\der^m h^{\rm TT}_{mn}=0$, $\eta^{mn}h^{\rm TT}_{mn}=0$. 
The linearized Einstein equations in this gauge are written as follows. 
\bea
 &&\brc{e^{-2\sgm}\Box_4-\der_y^2-4\dot{\sgm}\der_y}h^{\rm TT}_{mn} = 0, 
 \label{lin_eq1} \\
 &&\brc{e^{-2\sgm}\Box_4-\der_y^2-10\dot{\sgm}\der_y
 -4(\ddot{\sgm}+4\dot{\sgm}^2)}B  \nonumber\\
 &&\hspace{20mm}=\frac{4}{3}\kp^3\Re\brc{\frac{\der V}{\der\vph}(\vpcl)
 \tl{\vph}+\frac{1}{2}\sum_{\vtht^*=0,\pi}
 \frac{\der\lmd_{\vtht^*}}{\der\vph}(\vpcl)\tl{\vph}\cdot\dlt(y-R\vtht^*)}, 
 \label{lin_eq2} \\
 &&\brkt{\der_y+2\dot{\sgm}}B = -\frac{2}{3}\kp^3
 \Re\brkt{\dot{\bar{\vph}}_{\rm cl}\tl{\vph}}, 
 \label{lin_eq3}
\eea 
where $\Box_4\equiv \eta^{mn}\der_m\der_n$, $\kp\equiv 1/M_5$ 
and the dot denotes the derivative with respect to $y$. 
The first equation comes from the traceless part of $(m,n)$-component 
of the linearized Einstein equation, 
the second one from the trace part of $(m,n)$-component and 
the last one from $(m,y)$-component. 
The equation from $(y,y)$-component is not shown because 
it is not independent of the above equations. 
Here, we have used Eq.(\ref{bg_eqs}). 

From Eqs.(\ref{lin_eq1}) and (\ref{lin_eq2}), 
we can obtain the following boundary conditions. 
\be
 \der_y h^{\rm TT}_{mn}|_{y=y^*} = 0, \label{bd_cond1}
\ee
\be
 \left.\brc{\brkt{\der_y+2\dot{\sgm}}B+\frac{2}{3}\kp^3
 \Re\brkt{\frac{\der \lmd_{\vtht^*}}{\der\vph}(\vpcl)\tl{\vph}}}\right|_{y=y*} 
 = 0, \label{bd_cond2}
\ee
where $y^*\equiv R\vtht^*=0,\pi R$ are the locations 
of the boundaries.\footnote{
Strictly speaking, the boundaries cannot be expressed 
by the rigid value of $y$ in the Newton gauge \cite{montes}. 
However, we can express them by $y=y^*$ {\it at the linearized order}. 
(See Appendix~\ref{bd_cond_Ng}.)
} 
Using the jump condition~(\ref{jump_cond}), the second conditions are seen to
be equivalent to Eq.(\ref{lin_eq3}) and thus provide no new constraints. 

From Eqs.(\ref{lin_eq1}) and (\ref{bd_cond1}), 
we can see that $h^{\rm TT}_{mn}$ is 
decomposed into the 4D massless graviton and the massive K.K. gravitons 
\cite{EMS}. 
Since we are not interested in the 4D gravitational interactions, 
we will neglect $h^{\rm TT}_{mn}$ in the following. 

To simplify the discussion, we will assume that 
the background field configuration~$\vpcl(y)$ is real, 
and the fluctuation field~$\tl{\vph}$ satisfies the boundary condition: 
\be
 \tl{\vph}|_{y=y^*} = 0. \label{bd_cond_tlvph}
\ee 
This condition is realized, for example, 
in the Goldberger-Wise mechanism~\cite{GW} 
and its supersymmetric version~\cite{MO} considered in the next subsection. 
In this case, the boundary conditions~(\ref{bd_cond2}) become 
\be
 \brkt{\der_y+2\dot{\sgm}}B|_{y=y^*} = 0. \label{bd_cond_B}
\ee

Note that Eq.(\ref{lin_eq3}) indicates that the trace part of the metric 
perturbation~$B$ and the bulk scalar mode~$\vpR\equiv\Re\,\tl{\vph}$ 
describe the same physical degree of freedom. 
In fact, the linearized equation of motion for $\vpR$ can also be derived 
from Eqs.(\ref{lin_eq2}) and (\ref{lin_eq3}). 
Using Eq.(\ref{lin_eq3}), we can eliminate $\tl{\vph}$ 
in Eq.(\ref{lin_eq2}) and obtain an equation only for $B$. 
\be
 \brc{e^{-2\sgm}\Box_4-\der_y^2+2\brkt{
 \frac{\ddot{\vph}_{\rm cl}}{\dot{\vph}_{\rm cl}}-\dot{\sgm}}\der_y 
 +4\brkt{\dot{\sgm}\frac{\ddot{\vph}_{\rm cl}}{\dot{\vph}_{\rm cl}}
 -\ddot{\sgm}}}B=0. \label{eq_for_B}
\ee
We have used Eq.(\ref{bg_eqs}) again. 

From Eq.(\ref{eq_for_B}), the mode equation for $B$ is read off as 
\be
 \brc{-\der_y^2+2\brkt{\frac{\ddot{\vph}_{\rm cl}}{\dot{\vph}_{\rm cl}}
 -\dot{\sgm}}\der_y+4\brkt{\dot{\sgm}\frac{\ddot{\vph}_{\rm cl}}
 {\dot{\vph}_{\rm cl}}-\ddot{\sgm}}}f_{(p)}(y) 
 = m_{(p)}^2 e^{-2\sgm}f_{(p)}(y), \label{mode_eq}
\ee
where $m_{(p)}$ is the mass eigenvalue of the $p$-th K.K. mode. 
By solving Eq.(\ref{mode_eq}) with the boundary conditions~(\ref{bd_cond2}), 
we can decompose $B$ into 4D K.K. modes. 
\be
 B(x,y) = \sum_{p=0}^\infty f_{(p)}(y)b_{(p)}(x). 
\ee
The boundary condition at one boundary fixes the overall normalization 
of the mode functions~$f_{(p)}(y)$, and the condition at the other boundary 
determines the mass spectrum. 
As mentioned in Ref.~\cite{CGK}, 
the boundary conditions~(\ref{bd_cond_B}) ensure 
the orthogonality of the mode functions.

\subsection{Supersymmetric stabilization mechanism} \label{ex_stmch}
In the rest of this section, we will demonstrate the mode expansion 
and identify the radion mode 
in a model of Ref.~\cite{MO}. 
The radius stabilization mechanism in this model 
corresponds to a supersymmetric extension of 
the Goldberger-Wise mechanism~\cite{GW}. 

The stabilization sector consists of a hypermultiplet~$(H,H^C)$ with 
the bulk mass~$m$, where a chiral multiplet~$H$ ($H^C$) 
is defined as even (odd) under the orbifold parity. 
We will introduce the following superpotential, which provides source terms 
for the scalar component of $H^C$ on the boundaries.  
\be
 W_b \equiv \brc{J_0\dlt(y)-J_\pi\dlt(y-\pi R)}H,  
\ee
where $J_0$ and $J_\pi$ are constants. 
Here, we will make $J_0$ real by the phase redefinition of $H$.  

Now we will find a field configuration that preserves 
$\cN=1$ supersymmetry. 
The Killing spinor equations are 
\bea
 \dot{\sgm}+k\brc{1+\frac{\kp^3}{2}\brkt{\abs{h}^2+\abs{h^C}^2}}
 -\frac{1}{3}\kp^3m\brkt{\abs{h}^2-\abs{h^C}^2} \eql 0, \nonumber\\
 \der_y h+\brkt{m+\frac{3}{2}\dot{\sgm}}h \eql 0, \nonumber\\
 -\der_y h^C+\brkt{m-\frac{3}{2}\dot{\sgm}}h^C+J_0\dlt(y)-J_\pi\dlt(y-\pi R) 
 \eql 0, \nonumber\\
 mh^C h \eql 0, \label{eqs_MO}
\eea
where $k$ is a constant which becomes an AdS curvature in the limit 
of $J_0,J_\pi\to 0$.\footnote{
From the viewpoint of the superconformal gravity, 
$k$ and $m$ are determined by the gauge couplings 
for the graviphoton~\cite{AS}. 
(See Eqs.(\ref{rel_k-gc0}) and (\ref{rel_m-g0}).)
} 
The scalar fields~$h$ and $h^C$ are the scalar components 
of $H$ and $H^C$, respectively. 
The first equation comes from the supersymmetric variation of the gravitino, 
the second and third ones from the hyperinos, and the last one 
from the graviphotino.\footnote{
The graviphotino itself is unphysical after the superconformal gauge fixing, 
but the corresponding Killing spinor equation must be satisfied 
by the supersymmetric solution. } 

The source terms on the boundaries lead to the boundary conditions 
for $h^C$ as
\be
 \sbk{h^C}_0 = J_0, \;\;\;\;\; 
 \sbk{h^C}_{\pi} = J_\pi, \label{bd_cond_hc}
\ee
where the symbol~$\sbk{\cdots}_{\vtht^*}$ is defined by Eq.(\ref{sbk_bd}). 
Due to these conditions and the last equation in Eq.(\ref{eqs_MO}), 
only $h^C$ can have a nonzero background. 
Terms involving scalar fields in the first equation of Eq.(\ref{eqs_MO}) 
correspond to the backreaction of the scalar configuration on the metric. 
We will concentrate ourselves on the case that 
the backreaction is sufficiently small. 
Then, the classical background solution can be solved as 
\bea
 \sgm(y) \eql -ky-\frac{l^2}{24}e^{2\gm(y-\pi R)}+\cO(l^4), 
 \label{sol_sgm}\\
 h_{\rm cl}(y) \eql 0, \\
 h^C_{\rm cl}(y) \eql \frac{J_0}{2}e^{\gm y}\brkt{1+\cO(l^2)}, 
 \label{sol_hc}
\eea
where 
\be
 \gm \equiv m+\frac{3}{2}k, \label{def_c}
\ee
and $l\equiv\kp^{3/2}\abs{J_\pi}$ is a dimensionless parameter 
that parametrizes the size of the backreaction. 

Combining Eq.(\ref{sol_hc}) with Eq.(\ref{bd_cond_hc}), 
we can obtain the relation 
\be
 J_0=J_\pi e^{-\gm\pi R}. \label{rel_Js}
\ee
Thus, $J_\pi$ must be real for the above supersymmetric solution to exist. 
Eq.(\ref{rel_Js}) fixes the radius~$R$ to the definite value determine 
by the ratio of $J_0$ and $J_\pi$. 
Namely, the radius is stabilized. 
Note that the Killing spinor equations are the first-order differential 
equations and their solution contains only one integration constant. 
It is fixed by one of the boundary conditions~(\ref{bd_cond_hc}) 
and the other condition fixes the radius. 
However, the full equations of motion are the {\it second-order} differential 
equations and thus the most general solution contains 
two integration constants. 
In our case, the second integration constant is fixed 
by the minimization condition for the configuration energy 
(not just the stationary condition) 
because the preserved supersymmetry ensures 
the stability of the field configuration.  

Since only $h^C$ has the nonzero background configuration, 
we can apply the equations in the previous subsection 
by replacing $\vpcl(y)$ with $h^C_{\rm cl}(y)$. 
If we neglect the backreaction on the metric, 
the mode equation~(\ref{mode_eq}) can be easily solved. 
In this limit, the lightest mode~$b_{(0)}(x)$ is massless 
and its mode function is 
\be
 f_{(0)}(y) = C_{(0)}e^{-2\sgm(y)} = C_{(0)}e^{2ky}, 
 \label{f0_form}
\ee
where $C_{(0)}$ is a normalization constant. 
The other K.K. modes are expressed by the Bessel functions, 
and their mass spectrum is determined 
by the boundary condition~(\ref{bd_cond_B}) \cite{CGK}. 
If we take into account the backreaction, however, 
one can find that $b_{(0)}(x)$ obtains the following nonzero mass 
of order $\cO(l^2)$ as pointed out in Ref.~\cite{CGK}. 
\be
 m_{(0)}^2 = \frac{l^2k^2}{6}\brkt{1-\frac{2m}{k}}
 \brkt{\frac{3}{2}+\frac{m}{k}}^2
 e^{-2k\pi R}\frac{1-e^{-2k\pi R}}{1-e^{-(k-2m)\pi R}}+\cO(l^4). 
 \label{rad_mass}
\ee
The corresponding mode function satisfies the equation 
\be
 \der_y f_{(0)}+2\dot{\sgm}f_{(0)} = \cO(l^2). \label{ap_eq_f0}
\ee
Thus, we can find that the mode function of $b_{(0)}(x)$ 
in $\vpR\equiv\Re\,(h^C-h^C_{\rm cl})$ 
is suppressed by $\cO(l)$ factor from Eq.(\ref{lin_eq3}). 
In fact, $b_{(0)}(x)$ enters in $\vpR(x,y)$ as 
\bea
 \vpR \eql -l\frac{M_5^{\frac{3}{2}}\gm}{2k}e^{\gm(y-\pi R)}
 \brkt{e^{2ky}-1}
 \brc{1-e^{2(\gm-k)(\pi R-y)}\frac{(1-e^{-2k\pi R})(1-e^{-(k-2m)y})}
 {(1-e^{-2ky})(1-e^{-(k-2m)\pi R})}}b_{(0)}(x)  \nonumber\\
 &&+\cdots, \label{rel_vpR-b}
\eea
where the ellipsis denotes the massive K.K. modes. 
Note that this vanishes on the boundaries. 
In fact, the fluctuation field~$\tl{h}^C\equiv h^C-h^C_{\rm cl}$ satisfies 
the boundary condition~(\ref{bd_cond_tlvph}) due to Eq.(\ref{bd_cond_hc}). 
Thus, the mode functions satisfy the following orthonormal relation. 
\be
 \int_0^{\pi R}\dr y\; \frac{\brkt{3M_5^3m_{(p)}}^2}{2\dot{\vpcl}^2} 
 f_{(p)}(y)f_{(q)}(y) = \dlt_{pq}. \label{othnml_rel}
\ee
The constant factors $(3M_5^3m_{(p)})^2/2$ are determined 
by requiring the kinetic terms for $b_{(p)}(x)$ are canonically normalized.

\subsection{Radion mode} \label{rad_mode}
Now, we will identify the dynamical radion mode and its appearance 
in the action. 
Remember that the scalar fluctuation mode~$\vpR$ represents 
the same degree of freedom as the metric perturbation~$B$. 
Thus, each mode~$b_{(p)}(x)$ is contained not only in the metric 
but also in the bulk scalar~$h^C$. 
Therefore, the kinetic term for each mode comes from both 
the Einstein-Hilbert term and the kinetic term for $h^C$. 
From the Einstein-Hilbert term~$\cL_{\rm EH}$, 
the following term comes out. 
\bea
 \cL_{\rm EH} \eql \frac{M_5^3\sqrt{g}}{2}\cR \nonumber\\
 \eql 3M_5^3e^{2\sgm}\eta^{mn}\der_m B\der_n B+\cdots. 
\eea
We have performed the partial integral in the second equation. 

From the scalar kinetic term~$\cL_{\rm scalar}$, we will obtain 
\bea
 \cL_{\rm scalar} \eql \sqrt{g}\der^\mu\bar{h}^C\der_\mu h^C 
 = e^{2\sgm}\eta^{mn}\der_m\vpR\der_n\vpR+\cdots \nonumber\\
 \eql e^{2\sgm}\brc{-\frac{2}{3}\kp^3\dot{\vph}_{\rm cl}}^{-2}\eta^{mn}
 \der_m(\der_yB+2\dot{\sgm}B)\der_n(\der_yB+2\dot{\sgm}B)+\cdots. 
\eea
In the second equation, we have used Eq.(\ref{lin_eq3}). 

Therefore, the kinetic term for each mode in $B$ becomes 
\be
 \cL_{\rm kin} = M_5^3 e^{2\sgm}\sum_{p,q}
 \brc{3f_{(p)} f_{(q)} +\frac{9M_5^3}{4\dot{\vph}_{\rm cl}^2}
 (\der_y f_{(p)}+2\dot{\sgm}f_{(p)})(\der_y f_{(q)}
 +2\dot{\sgm}f_{(q)})}\eta^{mn}\der_mb_{(p)}\der_nb_{(q)}+\cdots  
 \label{kin_B}
\ee
The first term comes from $\cL_{\rm EH}$ 
and the second term from $\cL_{\rm scalar}$. 
First, let us consider the lightest mode~$b_{(0)}(x)$. 
Note that $\kp^3\dot{\vph}_{\rm cl}^2=\cO(l^2)$ and Eq.(\ref{ap_eq_f0}). 
Hence, the contribution from $\cL_{\rm scalar}$ is suppressed by $\cO(l^2)$ 
comparing to that of $\cL_{\rm EH}$. 
For the other modes, on the other hand, 
the first term in Eq.(\ref{kin_B}) is suppressed 
by $\cO(l^2)$ comparing to the second term 
because $\der_y f_{(p)}+2\dot{\sgm}f_{(p)}$ ($p\neq 0$) are not 
suppressed by $\cO(l^2)$. 
These facts suggest that $b_{(0)}(x)$ is mainly contained in the metric 
while the other modes are mainly in the bulk scalar~$\vpR$. 
After the $y$-integration and the partial integral, 
we can see that the kinetic terms for $b_{(p)}(x)$ are diagonal 
thanks to the mode equation~(\ref{mode_eq}) and 
the orthonormal relation~(\ref{othnml_rel}). 

Finally, we will identify the radion field. 
The radion field~$r(x)$ is defined as the proper length 
along the fifth dimension. 
Thus, at the linear order for the fields, 
\bea
 r(x) \defa \frac{1}{\pi}\int_0^{\pi R}\dr y\; 
 \abs{g_{yy}(x,y)}^{\frac{1}{2}} 
 = \frac{1}{\pi}\int_0^{\pi R}\dr y\; \brc{1-2B(x,y)} \nonumber\\
 \eql R-\frac{2}{\pi}\int_0^{\pi R}\dr y\;
 f_{(0)}(y)\cdot b_{(0)}(x)+\cO(l). 
\eea
The last term~$\cO(l)$ corresponds to the massive K.K. modes~$b_{(p)}$ 
($p\neq 0$). 
Therefore, the lightest mode~$b_{(0)}(x)$ can be identified with 
the radion fluctuation mode.

\subsection{Graviphoton mode} \label{Wy_mode}
When the model is embedded into 5D SUGRA, the theory contains 
the graviphoton~$W^0_\mu$. 
As is well-known, the radion supermultiplet involves $W^0_y$. 
Thus, we will consider the K.K. decomposition of $W^0_y$ 
in this subsection. 

The relevant terms in the lagrangian are \footnote{
Here, we will follow the notations of Ref.~\cite{KO1,KO2}. 
The graviphoton~$W^0_\mu$ is not canonically normalized there. 
}  
\be
 \cL = \sqrt{g}\sbk{-\frac{3}{8}M_5 F^{0\mu\nu}F^0_{\mu\nu}
 +\cD^\mu \bar{h}^C\cD_\mu h^C}+\cdots.  \label{L_W}
\ee
Here, $F^0_{\mu\nu}\equiv \der_\mu W^0_\nu-\der_\nu W^0_\mu$, 
and $\cD_\mu h^C$ is the covariant derivative for the graviphoton 
defined as 
\be
 \cD_\mu h^C \equiv \der_\mu h^C+i\kp\gm W^0_\mu h^C, 
\ee
where $\gm$ is defined in Eq.(\ref{def_c}). 
The ellipsis in Eq.(\ref{L_W}) denotes irrelevant terms to the discussion. 
Then, the linearized equations of motion for $W^0_\mu$ and $\vpI$ 
are~\footnote{
$\vpR$ is decoupled from $W^0_\mu$ at the linearized order.} 
\bea
 &&\hspace{-1cm}e^{-2\sgm}\brkt{\Box_4 W^0_m-\der_m\der_n W^{0\udl{n}}}
 -\der_y^2 W_m^0-2\dot{\sgm}\der_y W^0_m
 +\brkt{\der_y+2\dot{\sgm}}\der_m W^0_y  \nonumber\\
 &&\hspace{56mm}+\frac{8}{3}\kp^2\brc{\gm\vpcl\der_m\vpI
 +\kp\gm^2\vpcl^2 W^0_m} = 0, 
 \label{lin_eq_Wm} \\
 &&\hspace{-1cm}e^{-2\sgm}\Box_4 W^0_y+\frac{8}{3}\kp^3\gm^2\vpcl^2 W^0_y
 -e^{-2\sgm}\der_y\der_m W^{0\udl{m}}
 +\frac{8}{3}\kp^2\gm\brkt{\vpcl\der_y\vpI-\dot{\vph}_{\rm cl}\vpI} = 0, 
 \label{lin_eq_Wy} \\
 &&\hspace{-1cm}e^{-2\sgm}\Box_4\vpI-\der_y^2\vpI-4\dot{\sgm}\der_y\vpI
 +\brc{\gm(\gm-4k)-\frac{\kp^3\gm^2}{6}\vpcl^2}\vpI  \nonumber\\
 &&\hspace{1cm}+e^{-2\sgm}\kp\gm\vpcl\der_m W^{0\udl{m}}
 -\kp\gm\brc{\vpcl\der_y W_y^0+\brkt{2\dot{\vph}_{\rm cl}+4\dot{\sgm}\vpcl}
 W_y^0} = 0, \label{lin_eq_vpI}
\eea
where $\vpcl\equiv h^C_{\rm cl}$, $\vpI\equiv\Im\,\tl{h}^C$ 
and $W^{0\udl{m}}\equiv\eta^{mn}W^0_n$. 
Here, we have assumed that $\vev{W^0_y}=0$. 

In the limit of $l\to 0$, 
the gauge symmetry for the graviphoton is unbroken, 
and all K.K. modes of $W^0_y$ can be gauged away except for the zero-mode. 
On the other hand, $W^0_m$ has no zero-mode 
because of the orbifold projection. 
Thus, there is no common mode in $W^0_m$ and $W^0_y$ in this limit. 
Therefore, from Eq.(\ref{lin_eq_Wm}), we will obtain 
\be
 e^{-2\sgm}\brkt{\Box_4 W^0_m-\der_m\der_n W^{0\udl{n}}}
 -\der_y^2 W_m^0-2\dot{\sgm}\der_y W^0_m = 0. \label{Wm_cond1}
\ee
In the case that the stabilization sector exists, 
the K.K. modes of $W^0_y$ cannot be gauged away 
because the graviphoton gauge symmetry is broken by $\vpcl$. 
In this case, Eq.(\ref{Wm_cond1}) no longer holds. 
Instead, it is expected to be  
\be
 e^{-2\sgm}\brkt{\Box_4 W^0_m-\der_m\der_n W^{0\udl{n}}}
 -\der_y^2 W_m^0-2\dot{\sgm}\der_y W^0_m = \cO(l^2)  \label{Wm_cond2}
\ee
since Eq.(\ref{Wm_cond1}) recovers in the limit of $l\to 0$. 
Then, the following equation comes out from Eq.(\ref{lin_eq_Wm}). 
\be
 \brkt{\der_y+2\dot{\sgm}}W^0_y = -\frac{8}{3}\kp^2\gm\vpcl\vpI+\cO(l^2). 
 \label{const_WvpI}
\ee
This means that $W^0_y$ and $\vpI$ describe the same degree of freedom, 
just like $B$ and $\vpR$ in Eq.(\ref{lin_eq3}). 
We can show that Eq.(\ref{const_WvpI}) is consistent with 
Eqs.(\ref{lin_eq_Wy}) and (\ref{lin_eq_vpI}) at the leading order for $l$. 
This implies the validity of the expectation~(\ref{Wm_cond2}) 

Using Eq.(\ref{const_WvpI}), we can obtain the equation for only $W^0_y$ 
from Eq.(\ref{lin_eq_Wy}). 
\be
 \brc{e^{-2\sgm}\Box_4-\der_y^2+2\brkt{
 \frac{\dot{\vph}_{\rm cl}}{\vpcl}-\dot{\sgm}}\der_y
 +4\dot{\sgm}\frac{\dot{\vph}_{\rm cl}}{\vpcl}}W^0_y = \cO(l^2).  
 \label{eq_for_Wy}
\ee
Here, note that $\ddot{\sgm}$ and $\kp^3\vpcl^2$ are $\cO(l^2)$. 
For the supersymmetric solution~(\ref{sol_sgm})-(\ref{sol_hc}), 
this is the same equation as Eq.(\ref{eq_for_B}) 
at the leading order for $l$.
This suggests that the each modes in $B$ and $W^0_y$ belong to 
the same supermultiplet as expected. 
By similar discussion in Section~\ref{rad_mode}, 
we can see that the lightest mode is mainly contained in $W^0_y$, 
while the other K.K. modes are mainly in the bulk scalar~$\vpI$. 
Thus, the lightest mode of $W^0_y$ can approximately be expressed by 
the gauge-invariant Wilson line, 
\be
 w \equiv \frac{1}{\pi}\int_0^{\pi R}\dr y\; W^0_y. \label{def_C}
\ee

\section{Superspace action} \label{rad_sf_action}
In this section, we will derive the $\cN=1$ superspace action including 
the radion superfield. 
For this purpose, we will embed the radion field~$r(x)$ 
into the superspace action in an appropriate manner. 
Then, we can obtain the desired action by promoting the radion field 
to a chiral superfield. 

In the following, we will neglect the backreaction of the bulk scalar 
configuration on the metric because its effects are subdominant  
except for the radion mass. 
Namely, the warp factor is assumed as $\sgm(y)=-ky$ in this section. 
In Section~\ref{drv_radmass}, we will show that 
the correct radion mass~(\ref{rad_mass}) can also be obtained 
without including the backreaction by calculating the radion potential. 
Thus, neglecting the backreaction is a practical approximation. 

\subsection{Radion field in superspace action}
In order to express the action on the superspace, 
we will neglect 4D gravitational multiplet and its K.K. modes 
in the following. 
In the previous section, 
we have seen that the radion mode~$b(x)$~\footnote{
In contrast to fluctuation mode~$b_{(0)}(x)$ in the previous section, 
we will allow $b(x)$ to have a nonzero vacuum expectation value (VEV). } 
is contained only in the metric and its K.K. modes reside 
in the radius stabilizer field~$\vph$ in the limit of the small backreaction. 
Thus, the metric can be parametrized as~\footnote{
The off-diagonal components~$g_{my}$ can always be gauged away. }
\be
 ds^2 = e^{2F}\eta_{mn}dx^m dx^n-G^2 dy^2, \label{FG_metric}
\ee
where 
\be
 F = F(b(x),y), \;\;\;\;\;
 G = G(b(x),y). \label{def_FG}
\ee
Here, we will choose the coordinate~$y$ so that 
\be
 G(\vev{b},y) = 1, \label{ini_cond_G}
\ee
where $\vev{b}$ denotes VEV of $b(x)$. 
This is always possible by the redefinition of $y$. 
Then, the parameter~$R$, which indicates the range of $y$, 
is identical to the radius of the orbifold. 
Since the background spacetime is the slice of ${\rm AdS}_5$ 
with the curvature~$k$, 
\be
 F(\vev{b},y) = \sgm(y) = -ky. \label{ini_cond_F}
\ee

Although $\vev{b}$ is fixed to some definite value by the stabilization 
mechanism, $b(x)$ appears in the metric as if it is a modulus field 
since we neglect the backreaction. 
Hence, the bulk spacetime remains ${\rm AdS}_5$ with the curvature~$k$ 
when the background value of $b(x)$ is moved from the true value~$\vev{b}$. 
In order for the metric~(\ref{FG_metric}) to have this property, 
$F$ and $G$ must satisfy the following relation. 
%
\be
 G(b(x),y) = -\frac{1}{k}\der_y F(b(x),y). \label{G-Fp}
\ee
Note that both (\ref{naive_radion}) and (\ref{CGR_metric}) 
satisfy this condition. 

Next, we will embed the radion mode into the $\cN=1$ superspace. 
We will start with the superspace action derived 
in our previous paper~\cite{AS}.\footnote{
This action can also be obtained from that of Ref.~\cite{CST} 
by fixing the gravitational multiplet to its background value. } 
There, we derived 5D superspace action on the warped background 
directly from 5D SUGRA action. 
In order to incorporate the radion fluctuation mode~$b(x)$ 
into the action, 
we will replace the warp factor~$\sgm(y)$ and 
the background value of the f\"{u}nfbein~$\gey$ in Ref.~\cite{AS} 
with the $b$-dependent functions~$F$ and $G$, \ie, 
\bea
 \sgm(y) \to F(b(x),y), \nonumber\\
 \gey \to G(b(x),y). 
\eea

Basically, we will follow the notation of Ref.~\cite{AS}. 
We will introduce $\nV+1$ vector multiplets~$\cV^I$ ($I=0,1,\cdots,\nV$),
and $\nH+1$ hypermultiplets~$\cH^{\hat{\alp}}$ ($\hat{\alp}=0,1,\cdots,\nH$). 
The vector multiplet~$\cV^{I=0}$ denotes the graviphoton multiplet, 
and the hypermultiplet~$\cH^{\hat{\alp}=0}$ is 
the compensator multiplet.\footnote{
In this paper, we will consider the case of one compensator multiplet, 
for simplicity. 
An extension to the multi-compensator case is straightforward. 
} 
The remaining multiplets are physical ones. 
Here, we will use $\cH^{\hat{\alp}=1}$ as the radius stabilizer multiplet. 

From the vector multiplet~$\cV^I$, 
we can construct $\cN=1$ vector and chiral superfields~$V^I$ and $\Phi^I$. 
For simplicity, we will consider only abelian gauge groups 
in this paper. 
From the hypermultiplets~$\cH^{\hat{\alp}}$, 
we can construct a pair of chiral 
superfields~$(\Phi^{2\hat{\alp}+1},\Phi^{2\hat{\alp}+2})$. 
The explicit form of each superfield is collected 
in Appendix~\ref{def_sf}. 
The orbifold parity for each superfield is listed in Table.~\ref{Z2_parity}. 
\begin{table}[t]
\begin{center}
\begin{tabular}{|c||c|c|c|c||c|c|} \hline
\rule[-2mm]{0mm}{7mm} & $V^0$ & $\Phi_S^0$ & $V^{I\neq 0}$ & 
 $\Phi_S^{I\neq 0}$ & $\Phi^{2\hat{\alp}+1}$ & $\Phi^{2\hat{\alp}+2}$ 
 \\ \hline 
$Z_2$-parity & $-$ & $+$ & $+$ & $-$ & $-$ & $+$ \\ \hline
\end{tabular}
\end{center}
\caption{Orbifold parity for each superfield}
\label{Z2_parity}
\end{table}

Using these superfields, the 5D superconformal invariant action 
on the warped geometry can be written as follows. 
\bea
 S \eql \int\dr^5x\;\brkt{\cL_{\rm vector}+\cL_{\rm hyper}}, \nonumber\\
 \cL_{\rm vector} \eql \sbk{\int\dr^2\tht\;
 \frac{3C_{IJK}}{2}\brc{i\Phi_S^I\cW^J\cW^K
 +\frac{1}{12}\bar{D}^2\brkt{V^ID^\alp\der_yV^J-D^\alp V^I\der_yV^J}\cW^K_\alp}
 +\hc} \nonumber\\
 &&-e^{2F}\int\dr^4\tht\;G^{-2}C_{IJK}\cV_S^I\cV_S^J\cV_S^K, \nonumber\\ 
 \cL_{\rm hyper} \eql -2e^{2F}\int\dr^4\tht\;
 G\dmx\bar{\Phi}^\bt\brkt{e^{-2igV^It_I}}^\alp_{\;\;\gm}\Phi^\gm 
 \nonumber\\
 && -e^{3F}\sbk{\int\dr^2\tht\;
 \Phi^\alp\dmx\rho_{\bt\gm}\brkt{\der_y-2g\Phi_S^It_I}^\gm_{\;\;\dlt}\Phi^\dlt
 +\hc}, 
 \label{Sinv}
\eea
where 
\bea
 \cV_S^I \defa -\der_y V^I-i\Phi_S^I+i\bar{\Phi}_S^I, \nonumber\\
 \cW^I_\alp \defa -\frac{1}{4}\bar{D}^2D_\alp V^I 
\eea
are the gauge invariant quantities. 
$C_{IJK}$ is a real constant tensor which is completely symmetric 
for the indices, 
and $\dmx$ is a metric of the hyperscalar space and can be brought 
into the standard form~\cite{dewit} 
\be
 \dmx=\mtrx{\bdm{1}_2}{}{}{-\bdm{1}_{2\nH}}.  \label{def_dmx}
\ee
The first line of $\cL_{\rm vector}$ corresponds to the gauge kinetic terms 
and the supersymmetric Chern-Simons term. 
Here, all the directions of the gauging are chosen to $\sgm_3$-direction 
since the gauging along the other direction mixes $\Phi^{2\hat{\alp}+1}$ 
and $\Phi^{2\hat{\alp}+2}$, which have opposite $Z_2$-parities. 
Namely, the anti-hermitian generators~$t_I$ ($I=0,1,\cdots,\nV$) 
are assumed to be 
\be
 gt_0 = -i\dgnl{g^0_{\rm c}}{\bdm{g}^0}\otimes\sgm_3, \;\;\;\;\;
 gt_I = -i\dgnl{0}{\bdm{g}^I}\otimes\sgm_3, \;\;\;\;\;
 (I\neq 0)
\ee
where $\bdm{g}^I\equiv\diag(g_1^I,g_2^I,\cdots,g_{\nH}^I)$ 
$(I=0,1,\cdots,\nV)$ are $\nH\times\nH$ matrices of the gauge 
couplings for the physical hypermultiplets. 
We will also assume that the stabilizer multiplet~$\cH^{\hat{\alp}=1}$ 
is charged under only the graviphoton~$W^0_\mu$, \ie, $g_1^{I\neq 0}=0$.  
Otherwise, the gauge symmetries will be broken 
by the background configuration of the stabilizer field. 

In order to obtain the Poincar\'{e} SUGRA, we have to fix 
the extraneous superconformal symmetries 
by imposing the gauge fixing conditions, 
which are listed in Appendix~\ref{sc_gf}. 
After this gauge fixing, 
Eq.(\ref{Sinv}) reproduces the SUGRA action in Ref.~\cite{KO1,KO2} 
with the radion-dependent metric~(\ref{FG_metric}) 
if and only if $e^{2F}G$ is independent of 
the 4D coordinates~$x^m$.
Considering the conditions~(\ref{ini_cond_G}) and (\ref{ini_cond_F}), 
this condition can be written as 
\be
 2F+\ln G = 2\sgm. \label{FG_rel} 
\ee

Combining Eq.(\ref{FG_rel}) with Eq.(\ref{G-Fp}), 
we can determine $F$ and $G$ as 
\bea
 F \eql \frac{1}{2}\ln\brkt{e^{2\sgm}+\cI(b)}, \nonumber\\
 G \eql \frac{1}{1+e^{-2\sgm}\cI(b)}, 
\eea
where $\cI(b)$ is some function of only $b(x)$. 
From Eq.(\ref{ini_cond_G}), $\cI(b)$ satisfies 
\be
 \cI(\vev{b})=0. 
\ee
If we choose $\cI(b)$ as $\cI(b)=\tl{b}\equiv b-\vev{b}$, 
the metric~(\ref{FG_metric}) becomes that of Ref.~\cite{BNZ}. 
Changing the choice of $\cI(b)$ just corresponds 
to the field redefinition of $b(x)$, and causes no physical changes. 
Hence, we will take $\cI(b)=\tl{b}$ in the following. 
In this case, the fluctuation mode~$\tl{b}(x)$ is not canonically normalized. 
The relation to the normalized field~$b_{(0)}(x)$ is 
$\tl{b}(x)=2C_{(0)}b_{(0)}(x)$, 
where $C_{(0)}$ is the normalization constant in Eq.(\ref{f0_form}).\footnote{ 
Note that the metric~(\ref{FG_metric}) with the choice 
$\cI=2C_{(0)}b_{(0)}+\cO(b_{(0)}^2)$ 
is consistent with that in the Newton gauge at the linearized order, 
in which we worked in the previous section. }
The resulting forms of $F$ and $G$ are consistent with 
neglecting the K.K. graviton modes. 
In fact, as shown in Ref.~\cite{BNZ}, no additional terms are induced 
up to second order in spacetime derivatives 
after integrating out the K.K. gravitons. 

The radion field~$r(x)$ is related to $\tl{b}(x)$ as 
\bea
 r(x) \defa \frac{1}{\pi}\int_0^{\pi R}\dr y\;\abs{g_{yy}}^{\frac{1}{2}}
 =\frac{1}{\pi}\int_0^{\pi R}\frac{dy}{1+e^{-2\sgm}\tl{b}(x)}  \nonumber\\
 \eql R-\frac{1}{2k\pi}\ln\brkt{\frac{1+e^{2k\pi R}\tl{b}(x)}{1+\tl{b}(x)}}, 
 \label{r_to_b}
\eea
or equivalently, 
\be
 \tl{b}(x) = e^{-k\pi R}\frac{\sinh k\pi\brkt{R-r(x)}}{\sinh k\pi r(x)}. 
 \label{b_to_r}
\ee

In addition to the bulk action~(\ref{Sinv}), 
we can also construct the brane actions at the boundaries, 
which will be discussed in Section~\ref{brane_action}.

\subsection{Bulk action after gauge fixing}
To simplify the discussion, we will consider the maximally symmetric case 
in the vector sector, \ie,  
\be
 C_{IJK}M^IM^JM^K = 
 \brkt{M^{I=0}}^3-\frac{1}{2}M^{I=0}\sum_{J=1}^{\nV}\brkt{M^J}^2, 
\ee
where $M^I$ is a scalar component of the 5D vector multiplet~$\cV^I$. 
(See Appendix~\ref{sf_gf}.)
Then, after the superconformal gauge fixing, 
the lagrangian in Eq.(\ref{Sinv}) becomes 
\bea
 \cL \eql \brc{\int\dr^2\tht\;\frac{1}{4}G_{\rm c}\tl{\cW}^X\tl{\cW}^X+\hc}
 +e^{2\sgm}\int\dr^4\tht\;G^{-2}
 \brkt{\der_y\tl{V}^X+i\tl{\Phi}_S^X-i\bar{\tl{\Phi}}_S^X}^2 \nonumber\\
 &&-e^{2\sgm}\int\dr^4\tht\;
 (\bar{\Sgm}\Sgm)^{\frac{3}{2}}
 \brc{2M_5^3-G^{\frac{3}{2}}\brkt{\bar{H}e^{2\tlg^X\tl{V}^X}H
 +\bar{H}^Ce^{-2\tlg^X\tl{V}^X}H^C}}
 \nonumber\\
 &&+e^{3\sgm}\brc{\int\dr^2\tht\;\Sgm^3 H^C
 \brkt{\frac{1}{2}\lrder_y+\bdm{m}G_{\rm c}-2i\bdm{\tl{g}}^X
 \tl{\Phi}_S^X}H+\hc}  \nonumber\\ 
 &&+\cL_{\rm kin}^b,  
 \label{S_bulk}
\eea
where the summations for the indices~$X=1,2,\cdots,\nV$ are implicit, and 
\bea
 G \defa G(b(x),y) = \frac{1}{1+e^{-2\sgm(y)}\tl{b}(x)}, \label{def_G}\\
 G_{\rm c} \defa -2i\kp\vph_S^0 = G-i\kp W^0_y. 
 \label{def_Gc} 
\eea
Here, $\vph_S^0$ is a scalar component of $\Phi_S^0$. 
The derivative operator~$\lrder_y$ is defined as 
$A\lrder_y B \equiv A\der_y B-B\der_y A$. 
We have dropped quartic or higher order terms for the physical fields 
except for the radion because they are suppressed 
by the large Planck mass~$M_5$.  
Furthermore, we have assumed that the vacuum configuration 
preserves $\cN=1$ supersymmetry. 
The warp factor is determined by the Killing spinor condition 
from the supersymmetric variation of the gravitino, 
and as a result, the AdS curvature~$k$ can be written with respect to 
the compensator gauge coupling as~\footnote{
Note that we have neglected the backreaction of the scalar configuration.}  
\be
 k = \frac{2M_5 g^0_{\rm c}}{3}. \label{rel_k-gc0}
\ee

The physical superfields~$\tl{V}^X$, $\tl{\Phi}_S^X$ are defined as
\be
 \tl{V}^X = \sqrt{\frac{M_5}{2}}V^{I=X}, \;\;\;\;\;
 \tl{\Phi}_S^X = \sqrt{\frac{M_5}{2}}\Phi_S^{I=X}, 
\ee
and correspondingly, the physical (dimensionful) 
gauge couplings~$\bdm{\tl{g}}^X$ are defined as 
\be
 \bdm{\tl{g}}^X \equiv \sqrt{\frac{2}{M_5}}\bdm{g}^{I=X}. 
\ee

The mass matrix~$\bdm{m}$ is defined by 
\be
 \bdm{m} \equiv M_5\bdm{g}^0.  \label{rel_m-g0}
\ee

The compensator superfield~$\Sgm$ and 
the matter superfields~$H^v$, $H^{Cv}$ ($v=1,2,\cdots,\nH$) are defined as 
\bea
 \Sgm \defa \kp\brkt{\Phi^{\alp=2}}^{\frac{2}{3}} = 
 1-\tht^2\cF_\Sgm, \nonumber\\
 H^v \defa \frac{\sqrt{2}M_5^{\frac{3}{2}}G^{-\frac{3}{4}}
 \Phi^{\alp=2v+2}}{\Phi^{\alp=2}},  \nonumber\\
 H^{Cv} \defa \frac{\sqrt{2}M_5^{\frac{3}{2}}G^{-\frac{3}{4}}
 \Phi^{\alp=2v+1}}{\Phi^{\alp=2}}.
\eea
$\Phi^{\alp=1}$ and the fermionic component of $\Sgm$ 
only contribute $\cO(\kp)$ quartic or higher order terms, 
which are neglected in this paper. 
The $G$-dependence of each superfield is determined 
so that $G$ explicitly appears only through the form of $G_{\rm c}$ 
in the $d^2\tht$-integrals. 
This is because only $G_{\rm c}$ can be promoted to a chiral superfield 
in the $d^2\tht$-integrals. 
(See Eq.(\ref{def_GT}).) 

The radion kinetic term~$\cL^b_{\rm kin}$ comes from 
the cubic term for $\cV_S^0$ in Eq.(\ref{Sinv}), 
and is written as 
\be
 \cL_{\rm kin}^b = \frac{3}{4}e^{-2\sgm}M_5^3 G^2
 \eta^{mn}\der_m\tl{b}\der_n\tl{b}. 
\ee
In 5D SUGRA action, the same radion kinetic term comes from 
the Einstein-Hilbert term.

\subsection{Brane actions} \label{brane_action}
As well as the bulk action, we can express the brane actions 
localized on the boundaries in terms of the superfields~\cite{AS}. 
If we neglect fluctuations of 4D gravitational fields, 
the brane actions in Ref.~\cite{KO1} can be written as follows. 
\bea
 S_{\rm brane} \eql \sum_{\vtht^*=0,\pi}\int\dr^5x\; 
 c_{\vtht^*}\dlt(y-R\vtht^*)\cL^{(\vtht^*)}_{\rm brane}, \nonumber\\
 \cL^{(\vtht^*)}_{\rm brane} \eql \brc{\int\dr^2\tht\;
 f^{(\vtht^*)}_{AB}(S)\cW^A\cW^B+\hc}
 -e^{2F}\int\dr^4\tht\;\bar{\Sgm}\Sgm 
 \exp\brc{-K^{(\vtht^*)}(S,\bar{S},V)} \nonumber\\
 &&+e^{3F}\brc{\int\dr^2\tht\; \Sgm^3 P^{(\vtht^*)}(S)+\hc}, 
 \label{S_brane}
\eea
where $c_{\vtht^*}$ are some dimensionless constants 
which are assumed as small numbers. 
$f^{(\vtht^*)}_{AB}$, $K^{(\vtht^*)}$ and $P^{(\vtht^*)}$ are 
the (brane-localized) gauge kinetic functions, K\"{a}hler potentials, 
and superpotentials, respectively. 
The superfields~$V^A$ and $S^a$ are the vector and chiral superfields 
constructed from 4D superconformal multiplets. 
The explicit forms of them are collected in Appendix~\ref{4Dsf}. 
The indices~$A$ and $a$ run over not only the brane-localized multiplets 
but also induced ones on the boundaries from the bulk multiplets. 
For example, as chiral superfields on the boundary~$y=y^*$, 
\be
 S^v \equiv \left.M_5^{-\frac{1}{2}}G^{\frac{3}{4}}H^v\right|_{y=y^*} 
 \label{def_S}
\ee
can appear in the brane action~(\ref{S_brane}).\footnote{
$H^{Cv}$ are odd under the orbifold parity and vanish on the boundaries. }
The $G$-dependence is fixed by the requirement 
that the Weyl weights of $S^v$ must be zero~\cite{KU}. 
The warp factors in $\cL^{(\vtht^*)}_{\rm brane}$ 
come from the induced metric on the boundaries.

\subsection{Promotion to the radion superfield}
The radion fluctuation field~$\tl{b}(x)$ in $G$ and $G_{\rm c}$ 
should be promoted to a superfield 
because the unbroken $\cN=1$ supersymmetry exists. 
Note that there is still an ambiguity of the field redefinition of 
the radion field before promoting it to the superfield. 
As shown in Ref.~\cite{BNZ}, the proper length~$r(x)$ is suitable 
for the superfield description. 

Using Eq.(\ref{b_to_r}), we can rewrite $\cL^b_{\rm kin}$ 
in terms of $r(x)$ as follows. 
\be
 \cL^r_{\rm kin} = \frac{3}{16}M_5^3(k\pi)^2\brkt{1-e^{-2k\pi R}}^2
 \frac{e^{-2\sgm}G^2(r)}{\sinh^4 k\pi r}\eta^{mn}\der_m r\der_n r, 
 \label{L_r_kin}
\ee
where $G(r)\equiv G(\tl{b}(r))$. 

In the limit of the small backreaction (\ie, $l\to 0$), 
we have seen in Section~\ref{Wy_mode} 
that $W^0_y$ has only the zero-mode, which is described by 
the Wilson line~$w$ defined in Eq.(\ref{def_C}). 
By solving the equation of motion~\footnote{
Since $w$ is not contained in the bulk scalar~$\vph$ in the limit 
of $l\to 0$, we can drop the bulk scalar terms 
in the equation of motion. 
}, 
we can find that $w$ and $\tl{b}$ are contained in $W^0_\mu$ as~\cite{BNZ} 
\bea
 W^0_y \eql \frac{2k\pi e^{-2\sgm}}{e^{2k\pi R}-1}\cdot
 \frac{(1+\tl{b})(1+e^{2k\pi R}\tl{b})}
 {(1+e^{-2\sgm}\tl{b})^2}\cdot w,  \label{CinWy} \\
 W^0_m \eql \int_0^y\dr y^\prime\; 
 \der_m\brc{\frac{2k\pi e^{-2\sgm(y^\prime)}}{e^{2k\pi R}-1}\cdot
 \frac{(1+\tl{b})(1+e^{2k\pi R}\tl{b})}
 {(1+e^{-2\sgm(y^\prime)}\tl{b})^2}}w.  \label{CinWm}
\eea
Note that $W^0_m$ does not contain $w$ at the linearized order 
for the fluctuation fields, 
which is consistent with the analysis in Section~\ref{Wy_mode}. 


Substituting this expression into the kinetic term for $W^0_y$ 
in Eq.(\ref{L_W}) and rewriting $b(x)$ by $r(x)$, we can obtain 
\be
 \cL^w_{\rm kin} = \frac{3}{16}M_5(k\pi)^2(1-e^{-2k\pi R})^2
 \frac{e^{-2\sgm}G^2(r)}{\sinh^4 k\pi r}\eta^{mn}\der_m w\der_n w. 
 \label{L_C_kin}
\ee

Thus, if we define a 4D complex scalar~$\tau$ as 
\be
 \tau \equiv r+i\kp w, \label{def_T}
\ee
the above kinetic terms~$\cL^r_{\rm kin}$ and $\cL^w_{\rm kin}$ 
can be collected in the K\"{a}hler form. 
In order to see this explicitly,
let us derive the kinetic term for $\tau$ 
in 4D effective theory by performing the $y$-integral. 
\be
 \cL^{(4)}_{\rm kin} = \frac{3M_5^3k\pi^2}{8}
 \frac{1-e^{-2k\pi R}}{\sinh^2 k\pi\Re\,\tau}
 \eta^{mn}\der_m\bar{\tau}\der_n\tau. 
\ee
This is certainly the K\"{a}hler form with the K\"{a}hler potential 
\be
 K^{\rm eff}_{\rm rad}(\tau,\bar{\tau}) = -3M_{\rm P}^2
 \ln\brkt{1-e^{-k\pi(\tau+\bar{\tau})}}. 
 \label{K_rad}
\ee
Here, $M_{\rm P}\equiv \brkt{M_5^3(1-e^{-2k\pi R})/(2k)}^{1/2}$ 
is the 4D effective Planck mass. 
This suggests that the appropriate definition of 
the radion field for the promotion to a chiral superfield 
is the proper length~$r(x)$,\footnote{
For other definitions of the radion field, for example $b(x)$, 
the kinetic term cannot be written as the K\"{a}hler form.}
and all $r(x)$ in the chiral superspace should be associated with $w(x)$ 
in the form of $\tau$. 
Namely, $G_{\rm c}$ defined in Eq.(\ref{def_Gc}) should be understood as 
\be
 G_{\rm c}(r,w) = G(\tau) = \brc{1+e^{-2\sgm(y)}e^{-k\pi R}
 \frac{\sinh k\pi(R-\tau)}{\sinh k\pi\tau}}^{-1}. 
\ee
Here, we have used the relation~(\ref{b_to_r}). 
Then, from Eq.(\ref{def_Gc}), $G$ and $W^0_y$ are identified as 
\be
 G \equiv \Re\,G(\tau), \;\;\;\;\; 
 W^0_y \equiv -M_5\Im\,G(\tau). \label{new_def_GW}
\ee
These are identical to the expressions in Eqs.(\ref{def_G}) and (\ref{CinWy}) 
up to the linear order for $w$, 
but deviate from them beyond the linear order for $w$. 
This is not a problem because Eqs.(\ref{def_G}) and (\ref{CinWy}) 
are solutions of the equations of motion {\it up to the linear order} for $w$. 
In fact, we have implicitly assumed that the metric is independent of $w$, 
but this is only valid at the linearized level. 
Beyond the linearized order, the Einstein equation involves $W^0_y$ 
and the metric also has the $w$-dependence. 
Thus, Eq.(\ref{new_def_GW}) provides the modified expressions 
of $G$ and $W^0_y$ which are valid at all orders for $r$ and $w$.  

The appearance of $w$ in Eq.(\ref{CinWm}) 
induces nonvanishing contributions to the action from $F^0_{mn}$. 
In Refs.~\cite{BNZ,BR}, such contributions are ignored 
because they are of higher order in the derivative expansion. 
In our superspace formalism, on the other hand, 
$W^0_m$ is dropped from the beginning 
because it cannot be incorporated into the superspace. 
Thus, no such higher derivative terms appear in our superspace action. 

Now we will promote the complex scalar field~$\tau(x)$ 
to a chiral superfield~$T(x,\tht)$. 
Namely, $G_{\rm c}$ and $G$ in Eq.(\ref{S_bulk}) are promoted as 
\bea
 G_{\rm c} & \to & G(T) = \brc{1+e^{2ky}e^{-k\pi R}
 \frac{\sinh k\pi(R-T)}{\sinh k\pi T}}^{-1}, \nonumber\\
 G & \to & G_{\rm R} \equiv \Re\,G(T).  \label{def_GT}
\eea
As a result, the bulk lagrangian becomes 
\bea
 \cL \eql \brc{\int\dr^2\tht\;\frac{1}{4}G(T)\tl{\cW}^X\tl{\cW}^X+\hc}
 +e^{2\sgm}\int\dr^4\tht\;G_{\rm R}^{-2}
 \brkt{\der_y\tl{V}^X+i\tl{\Phi}_S^X-i\bar{\tl{\Phi}}_S^X}^2 \nonumber\\
 &&+e^{2\sgm}\int\dr^4\tht\;
 G_{\rm R}^{\frac{3}{2}}\brkt{\bar{H}e^{2\tlg^X\tl{V}^X}H
 +\bar{H}^Ce^{-2\tlg^X\tl{V}^X}H^C}
 \nonumber\\
 &&+e^{3\sgm}\brc{\int\dr^2\tht\; H^C
 \brkt{\frac{1}{2}\lrder_y+\bdm{m}G(T)-2i\bdm{\tl{g}}^X\tl{\Phi}_S^X}H+\hc} 
 \nonumber\\ 
 &&+e^{2\sgm}\int\dr^4\tht\; K_{\rm rad}(T,\bar{T}),  \label{fnl_L_bulk}
\eea
where the radion K\"{a}hler potential~$K_{\rm rad}(T,\bar{T})$ is 
\be
 K_{\rm rad}(T,\bar{T}) = -3M_5^3\ln G_{\rm R}. \label{Krad5D}
\ee
Since we assume the supersymmetric stabilization mechanism, 
$\cF_\Sgm$ does not have a non-zero VEV. 
Hence, from now on, we will drop the $\Sgm$-dependence. 
The 4D radion K\"{a}hler potential~$K_{\rm rad}^{(4)}(T,\bar{T})$ 
in the effective theory is obtained by the $y$-integration. 
\be
 K_{\rm rad}^{(4)}(T,\bar{T}) \equiv \int_0^{\pi R}\dr y\; 
 e^{2\sgm}K_{\rm rad}(T,\bar{T}).  \label{def_Krad4D}
\ee
We will not show its explicit expression here because it is lengthy 
and complicated. 
Although $K_{\rm rad}^{(4)}(T,\bar{T})$ has a different form 
from $K_{\rm rad}^{\rm eff}(T,\bar{T})$ in Eq.(\ref{K_rad}), 
the corresponding K\"{a}hler metric is identical to 
that of $K^{\rm eff}_{\rm rad}(T,\bar{T})$ 
up to the linear order for $\Im\,T$. 
\be
 \frac{\der^2}{\der T\der\bar{T}}K^{(4)}_{\rm rad}(T,\bar{T}) 
 = \frac{\der^2}{\der T\der\bar{T}}K^{\rm eff}_{\rm rad}(T,\bar{T})
 +\cO\brkt{(\Im\,T)^2}. 
\ee

Since the zero-modes of the gauge superfields~$\tl{V}^X$ 
have constant mode functions, 
the radion couplings to the superfield strengths of 
the zero-modes~$\tl{\cW}_{(0)}^X$ 
in 4D effective lagrangian becomes simple. 
\be
 \cL^{(4)}_{\rm gauge} = \int_0^{\pi R}\dr y\brc{\int\dr^2\tht\; 
 \frac{1}{4}G(T)\tl{\cW}^X\tl{\cW}^X+\hc}
 = \brc{\int\dr^2\tht\; \frac{\pi}{4}T\tl{\cW}_{(0)}^X\tl{\cW}_{(0)}^X+\hc}
 +\cdots, 
\ee
where the ellipsis denotes terms involving the massive K.K. modes. 
This coincides with the radion couplings of Ref.~\cite{LS}. 
On the other hand, the couplings to the other K.K. modes 
have complicated forms due to their nontrivial mode functions. 

Next, we will derive the boundary superspace lagrangians. 
The radion field~$r(x)$ contained in $F$ in Eq.(\ref{S_brane}) 
should be promoted to the superfield~$T$. 
The resulting boundary lagrangians are 
\bea
 \cL^{(\vtht^*)}_{\rm brane} \eql \brc{\int\dr^2\tht\; 
 f^{(\vtht^*)}_{AB}(S)\cW^A\cW^B+\hc}
 -e^{2\sgm(y^*)}\int\dr^4\tht\;\frac{1}{\Re\,G_{\vtht^*}(T)}
 \exp\brc{-K^{(\vtht^*)}(S,\bar{S},V)} \nonumber\\
 &&+e^{3\sgm(y^*)}\brc{\int\dr^2\tht\; 
 G_{\vtht^*}^{-\frac{3}{2}}(T)P^{(\vtht^*)}(S)+\hc}, 
 \label{S_brane2}
\eea
where $\vtht^*=0,\pi$, and  
\bea
 G_0(T) \defa G(T)|_{y=0} = \frac{1-e^{-2k\pi T}}{1-e^{-2k\pi R}}, \nonumber\\
 G_\pi(T) \defa G(T)|_{y=\pi R} = \frac{e^{2k\pi T}-1}{e^{2k\pi R}-1}. 
\eea

In the $d^2\tht$-integral, this is the only way of promoting $G(r)$ 
because of the holomorphicity. 
In the $d^4\tht$-integral, on the other hand, 
it seems that there is an ambiguity in the promotion. 
In fact, $G(r)$ might be promoted to $G(\Re\,T)$ 
instead of $\Re\,G(T)$. 
This ambiguity is removed by the requirement that 
the superspace action reproduces 
the brane-localized radion kinetic terms 
which originate from the 4D Einstein-Hilbert terms for the induced metric. 
Only the promotion applied in Eq.(\ref{S_brane2}) reproduces those terms. 
After the promotion of $r(x)$, note that 
the definition of $S^v$ in Eq.(\ref{def_S}) should be modified as 
\be
 S^v \equiv \left.M_5^{-\frac{1}{2}}G^{\frac{3}{4}}(T)H^v\right|_{y=y^*} 
\ee
because of the holomorphicity.  

Here, note that Eq.(\ref{S_brane2}) includes the Wilson line~$w$ 
through the $T$-dependence. 
This mode is originally contained in $\Phi_S^0$ in Eq.(\ref{Sinv}). 
On the other hand, the superconformal multiplets corresponding to $\Phi_S^I$ 
in our notation can be absent in the brane actions in Ref.~\cite{KO1}. 
(See Eq.(5.1) in Ref.~\cite{KO1}.)
The way of introducing $\Phi_S^I$ in the brane action 
is not mentioned in Ref.~\cite{KO1} just because it is a hard task to find 
how they appear in the 4D action formulae 
in a 5D superconformal-invariant way due to their nontrivial transformation 
properties. 
Namely, the possibility of the appearance of $\Phi_S^I$ in the brane action 
is not excluded yet. 
The appearance of $w$ in Eq.(\ref{S_brane2}) suggests 
the appearance of $\Phi_S^0$ in the boundary action in Ref.~\cite{KO1}. 
At least, from our superspace approach, it is inevitable for 
$w$ to appear in the boundary actions 
due to the existence of the radion mode in the induced metric. 

Finally, we will comment on the explicit forms of 
the components in each superfield listed in Appendix~\ref{sf_gf} 
after the radion~$r$ is promoted to the superfield~$T$. 
After the promotion, the expressions in Appendix~\ref{sf_gf} 
receive some modifications. 
They will newly obtain the dependence of $w$ 
and the zero-mode of the gravitino~$\psi_y$.  
However, such explicit expressions are irrelevant to the discussions 
once the superfield description is completed.

\subsection{Radion mass} \label{drv_radmass}
In Section~\ref{ex_stmch}, we have calculated the non-zero radion mass 
by including the backreaction of the scalar configuration on the metric. 
Here, we will show that we can also obtain the correct radion mass 
from the radion potential without including the backreaction. 

First, we will redefine the superfields as follows. 
\bea
 H & \to & G^{-\frac{m}{2k}}(T)H, \nonumber\\
 H^C & \to & G^{\frac{m}{2k}}(T)H^C. 
 \label{Hs_redef}
\eea
Then, the superspace lagrangian of the model discussed 
in Section~\ref{ex_stmch} can be written as 
\bea
 \cL \eql e^{2\sgm}\int\dr^4\tht\; 
  G_{\rm R}^{\frac{3}{2}}\brkt{\abs{G^{-\frac{m}{2k}}(T)H}^2
  +\abs{G^{\frac{m}{2k}}(T){H}^C}^2} \nonumber\\
 &&+e^{3\sgm}\brc{\int\dr^2\tht\; H^C\brkt{
  \frac{1}{2}\lrder_y+m}H+\hc}+e^{2\sgm}\int\dr^4\tht\;K_{\rm rad}(T,\bar{T}) 
 \nonumber\\
 &&+\dlt(y)\cL^{(0)}+\dlt(y-\pi R)\cL^{(\pi)}, 
 \label{L_MO}
\eea
where $K_{\rm rad}$ is defined in Eq.(\ref{Krad5D}), 
and the boundary lagrangians are 
\be
 \cL^{(\vtht^*)} = \left.e^{3\sgm}\brc{\int\dr^2\tht\; 
 G^{-\brkt{\frac{3}{4}+\frac{m}{2k}}}(T) e^{i\vtht^*}J_{\vtht^*}H
 +\hc}\right|_{y=y^*}.  \label{MO_bd_terms}
\ee
The brane-localized K\"{a}hler potentials are not introduced. 

Note that the $T$-dependence disappears from the $d^2\tht$-integration 
in the bulk action by the redefinition~(\ref{Hs_redef}). 
This makes it easier to calculate the radion potential. 
A naive way of deriving the low-energy effective theory is simply dropping 
the massive K.K. modes from the beginning and performing the $y$-integral. 
However, the effective theory obtained by such `zero-mode truncation' 
may receive some corrections in the process of 
integrating out the massive K.K. modes. 
The massive modes are integrated out by using their equations of motion. 
If a linear term for a massive K.K. mode includes a light mode 
like the radion~$T$, the procedure of integrating out such a mode 
induces a correction to the effective potential~\cite{HLW}. 
In Eq.(\ref{L_MO}), however, note that such $T$-dependent linear terms 
for the massive K.K. modes of $H$ or $H^C$ appears 
only in the boundary terms except for the K\"{a}hler terms.\footnote{
The $T$-dependence in the kinetic terms of the massive modes 
is harmless because its effect is suppressed in low energies. } 
Since $H^C$ vanishes on the boundaries, the above-mentioned corrections 
to the radion potential are not induced in our case.\footnote{
The procedure of integrating out $H_{(p)}$ ($p\neq0$) does not induce extra 
$T$-dependence. } 
Therefore, the naive zero-mode truncation can be used 
to calculate the radion potential. 

By dropping the massive K.K. modes, the superfields~$H$ and $H^C$ become  
\bea
 H(x,y,\tht) \eql C^H_{(0)}e^{\brkt{\frac{3}{2}k-m}y}\cdot H_{(0)}(x,\tht), 
 \nonumber\\
 H^C(x,y,\tht) \eql h^C_{\rm cl}(y) = \frac{J_0}{2}e^{\gm y}, 
\eea 
where $C^H_{(0)}$ is a normalization factor of the mode function, 
and $\gm\equiv \frac{3}{2}k+m$. 
As mentioned in Section~\ref{rad_mode}, the zero-mode of $H^C$ 
is suppressed by an $\cO(l)$ factor, and can be neglected. 

Then, the effective 4D lagrangian is obtained by the $y$-integral as 
\be
 \cL^{(4)} = \int\dr^4\tht\; \cK^{(4)}
 +\brc{\int\dr^2\tht\; P^{(4)}+\hc}, 
\ee
where 
\bea
 \cK^{(4)} \eql K^{(4)}_{\rm rad}(T,\bar{T}) 
 +\int_0^{\pi R}\dr y\; 
 e^{(k+2m)y}G_{\rm R}^{\frac{3}{2}}\abs{\frac{J_0}{2}G^{\frac{m}{2k}}(T)}^2, 
 \nonumber\\
 &&+ \int_0^{\pi R}\dr y\; 
 e^{(k-2m)y}G_{\rm R}^{\frac{3}{2}}
 \abs{G^{-\frac{m}{2k}}(T)C^H_{(0)}H_{(0)}}^2  \nonumber\\
 P^{(4)} \eql \frac{1}{2}G^{-\frac{\gm}{2k}}_0(T)
 \brkt{J_0-J_\pi e^{-\gm\pi T}}C^H_{(0)}H_{(0)}. 
\eea
Here, $K^{(4)}_{\rm rad}$ is defined in Eq.(\ref{def_Krad4D}). 
The superpotential~$P^{(4)}$ originates only 
from the boundary terms~(\ref{MO_bd_terms}). 
Note that the factor~$\frac{1}{2}$ in $P^{(4)}$ comes from 
the $y$-integral of the $\dlt$-functions 
since the integral interval is taken as $[0,\pi R]$. 

The scalar potential~$V_{\rm scalar}$ is calculated from this as 
\be
 V_{\rm scalar} = \brkt{g_K^{-1}}^{a\bar{b}}
 P^{(4)}_a\bar{P}^{(4)}_{\bar{b}}, 
\ee
where $a,b=\tau,h_{(0)}$ ($h_{(0)}$ is a scalar component of $H_{(0)}$), 
$P^{(4)}_a\equiv \der P^{(4)}/\der a, \cdots$ 
and $(g_K)_{a\bar{b}}\equiv \cK^{(4)}_{a\bar{b}}$ is the K\"{a}hler metric. 
We have dropped contributions from $\cF_\Sgm$ 
since they are irrelevant to the following discussion. 
The minimization conditions of this potential lead to the following vacuum. 
\bea
 \vev{h_{(0)}} \eql 0, \nonumber\\
 J_0-J_\pi e^{-\gm\pi\vev{\tau}} \eql 0. 
\eea
Thus, the radius~$\vev{r}=\Re\,\vev{\tau}$ 
is certainly stabilized to a finite value. 
The second equation reduces to Eq.(\ref{rel_Js}) 
since $\vev{r}=R$. 

For the calculation of the radion mass, we can restrict the potential 
to the section of $h_{(0)}=0$. 
Then, the radion potential~$V_{\rm rad}$ reduces to the following 
simple form. 
\bea
 V_{\rm rad}(\tau,\bar{\tau}) \eql 
 \left.\brkt{\cK^{(4)}_{H_{(0)}\bar{H}_{(0)}}}^{-1}
 \abs{P^{(4)}_{H_{(0)}}}^2\right|_{h_{(0)}=0} \nonumber\\
 \eql \frac{\abs{\frac{1}{2}G_0^{-\frac{\gm}{2k}}(\tau)}^2}
 {\int_0^{\pi R}\dr y\;e^{(k-2m)y}G_{\rm R}^{\frac{3}{2}}
 \abs{G^{-\frac{m}{2k}}(\tau)}^2}\cdot
 \abs{J_0-J_\pi e^{-\gm\pi\tau}}^2+\cO(l^4), 
\eea
where $l\equiv \kp^{3/2}\abs{J_\pi}$. 

Considering canonical normalization of the radion kinetic term, 
we can calculate the radion mass as 
\bea
 m_{\rm rad}^2 \eql \left.\brkt{\cK^{(4)}_{T\bar{T}}}^{-1}
 \frac{\der^2 V_{\rm rad}}{\der\tau\der\bar{\tau}}\right|_{\tau=R} \nonumber\\
 \eql \frac{l^2k^2}{6}\brkt{1-\frac{2m}{k}}
 \brkt{\frac{3}{2}+\frac{m}{k}}^2 e^{-2k\pi R}
 \frac{1-e^{-2k\pi R}}{1-e^{-(k-2m)\pi R}}+\cO(l^4). 
 \label{m_rad}
\eea
This radion mass is exactly identical to Eq.(\ref{rad_mass}) 
that is obtained by solving the mode equation. 
This supports the validity of the $T$-dependence of the action 
obtained in the previous subsection. 
In Ref.~\cite{MO}, the radion mass is calculated by using 
the superspace action of Ref.~\cite{MP}, which is based on 
the naive ansatz~(\ref{naive_radion}).  
Their result is similar to ours if $kR\simgt 1$.\footnote{
The radion mass calculated in Ref.~\cite{MO} is factor two 
larger than our result besides an extra factor~$(1-e^{-2k\pi R})$, 
but we think that this factor two is just their calculation error. }
(For example, $kR\simeq 12$ in the original Randall-Sundrum model.)

In the above derivation of the radion mass, the backreaction 
on the warp factor contributes only 
to a higher order correction of $\cO(l^4)$, 
in contrast to the derivation in Section~\ref{ex_stmch}. 
Now, the leading $\cO(l^2)$ contribution comes 
from the scalar configuration~$h^C_{\rm cl}(y)$, not from the backreaction. 
Therefore, we can obtain the correct radion mass without 
including the backreaction term in the warp factor.

\section{Summary} \label{summary}
We have derived 5D superspace action 
including the {\it dynamical} radion superfield, 
and clarified its couplings to the bulk and the boundary matter superfields. 
The resulting superspace lagrangian is provided by 
Eqs.(\ref{fnl_L_bulk}) and (\ref{S_brane2}) with Eq.(\ref{def_GT}). 
Our result is obtained in a systematic way 
based on the superconformal formulation of 5D SUGRA in Ref.~\cite{KO1,KO2}. 

The $T$-dependence of our action is different from 
that of Ref.~\cite{MP}, 
which is based on the naive ansatz~(\ref{naive_radion}). 
As we mentioned at the end of Section~\ref{drv_radmass}, 
the correct order of the radion mass can also be obtained 
by using the action of Ref.~\cite{MP} if $kR\simgt 1$, 
although it slightly deviates from the correct value~(\ref{rad_mass})  
by a factor~$(1-e^{-2k\pi R})$. 
This is because the radion potential 
reflects only infrared behavior of the radion field~$r(x)$,  
and the radion field defined in Ref.~\cite{MP} has a common 
infrared behavior with ours. 
(VEV of either radion field corresponds to the radius of the orbifold. ) 
However, when the radion field is treated 
as a dynamical degree of freedom, the difference of its definition 
becomes relevant. 
For example, the radion couplings to the K.K. modes of the matter fields 
are quite different from those of Ref.~\cite{MP}. 
In that case, calculations should be performed by using our result. 

Note that $K_{\rm rad}^{(4)}(T,\bar{T})$ defined in Eq.(\ref{def_Krad4D}) 
is different from $K_{\rm rad}^{\rm eff}(T,\bar{T})$ in Eq.(\ref{K_rad}), 
which is derived in Refs.\cite{BNZ,BR}. 
This difference cannot be removed by the redefinition of the superfields. 
The most general form of $G(T)$ in Eq.(\ref{def_GT}) after the redefinition 
of $T$ is written in the form 
\be
 G(T) = \frac{1}{1+e^{2ky}f(T)}, 
\ee
where $f(T)$ is a holomorphic function of $T$ and satisfies 
$\vev{f(T)}=0$. 
Thus we can calculate the most general form of $K_{\rm rad}^{(4)}(T,\bar{T})$ 
by substituting the above expression into Eqs.(\ref{Krad5D}) and 
(\ref{def_Krad4D}) and performing the $y$-integration. 
Then, we can easily see that $K_{\rm rad}^{(4)}(T,\bar{T})$ cannot be 
reduced to $K_{\rm rad}^{\rm eff}(T,\bar{T})$ in Eq.(\ref{K_rad})
except for the flat case, 
no matter how we choose the function~$f(T)$. 
Only in the flat case, the former is reduced to the latter 
without any redefinition of $T$. 
We will comment on this case at the end of this section. 

In Ref.~\cite{BR}, it is concluded that 
the K\"{a}hler potential is a function of only $\Re\,\tau$ 
from the invariance of the bosonic part of the action 
under a constant shift of $W^0_y$. 
However, we should emphasize that 
the constant shift of $W^0_y$ does not correspond 
to a constant shift of its zero-mode~$w$. 
(See Eq.(\ref{CinWy}), for example.)
Note that $w$ in Eq.(\ref{def_T}) must be a {\it fluctuation mode}. 
If it has a nonzero VEV, $G$ and $W^0_y$ defined by Eq.(\ref{new_def_GW}) 
deviate from Eqs.(\ref{def_G}) and (\ref{CinWy}) even at the leading order 
for the fluctuation modes. 
Thus, the constant shift of $w$ is not allowed. 
In the pure supergravity case discussed in Ref.~\cite{BNZ}, 
the bosonic part of the action is certainly invariant 
under the constant shift of $W^0_y$. 
However, such constant shift affects only VEV of $W^0_y$ 
and not the fluctuation mode~$w$. 
Recall that we have assumed $\vev{W^0_y}=0$ in Eq.(\ref{def_C}). 
If we admit a nonzero $\vev{W^0_y}$, Eq.(\ref{def_C}) should be written as 
\be
 w \equiv \frac{1}{\pi}\int_0^{\pi R}\dr y\; \brkt{W^0_y-\vev{W^0_y}}. 
\ee
Therefore, we conclude that the `mode' which should not appear 
in $K_{\rm rad}^{(4)}(\tau,\bar{\tau})$ is not $w=M_5\Im\,\tau$, 
but $W_y^0=-M_5\Im\,G(\tau)$. 

In the last section of Ref.~\cite{BNZ}, 
another derivation of the radion K\"{a}hler 
potential~$K_{\rm rad}^{\rm eff}(T,\bar{T})$ is presented, 
which is based on the assumption 
that 4D effective theory is described by 4D Einstein supergravity. 
However, when the radion is dynamical, the gravity deviates from 
the ordinary 4D Einstein gravity since the radion behaves like 
a Brans-Dicke scalar~\cite{CGRT,GT}.\footnote{
Of course, the Einstein gravity will recover below the radion mass scale.} 
Therefore, the discussion there may not be applicable 
to the derivation of the radion K\"{a}hler potential. 
Thus, the deviation of our result from that of Ref.~\cite{BNZ} 
does not lead to an immediate contradiction. 

In this paper, we have assumed the supersymmetric radius stabilization, 
and dropped the dependence of the compensator superfield. 
Actually, the $F$-terms of the compensator and the radion superfields 
are closely related to each other, 
and the promotion of the radion to the superfield involves 
a modification of $\cF_\Sgm$. 
This fact becomes relevant when we consider the Scherk-Schwarz breaking 
of the supersymmetry~\cite{SS}. 
We will discuss this issue in the subsequent paper. 

Finally, we will comment on the flat limit~(\ie, $k\to 0$). 
In this limit, $G(T)$ becomes independent of $y$ and reduces to a simple form, 
\be
 G(T) = \frac{T}{R}. 
\ee
Then, the $T$-dependence of the action becomes greatly simplified. 
Furthermore, from Eqs.(\ref{Krad5D}) and (\ref{def_Krad4D}), 
the 4D radion K\"{a}hler potential becomes the following no-scale form 
up to a constant. 
\be
 K_{\rm rad}^{(4)}(T,\bar{T}) = -3M_{\rm P}^2\ln\brkt{T+\bar{T}}, 
\ee
where $M_{\rm P}=\brkt{\pi RM_5^3}^{1/2}$ is the 4D Planck mass.

\vspace{3mm}

\begin{center}
{\bf Acknowledgments}
\end{center}
The authors would like to thank Minoru Eto 
for useful information. 
H.~A. is supported by KRF PBRG 2002-070-C00022. 
Y.~S. is supported from the Astrophysical Research Center 
for the Structure and Evolution of the Cosmos (ARCSEC) 
funded by the Korea Science and Engineering Foundation 
and the Korean Ministry of Science. 

\appendix

\section{Equations of motion for the background} \label{EOM_bg}
The background~(\ref{bkgd}) satisfies the following equations. 
\bea
 \dot{\sgm}^2 \eql \frac{\kp^3}{6}\brc{\abs{\dot{\vph}_{\rm cl}}^2
 -V(\vpcl)}, \nonumber\\
 \ddot{\sgm} \eql -\frac{\kp^3}{3}\brc{2\abs{\dot{\vph}_{\rm cl}}^2
 +\sum_{\vtht^*=0,\pi}\lmd_{\vtht^*}(\vpcl)\dlt(y-R\vtht^*)}, \nonumber\\
 \ddot{\vph}_{\rm cl}+4\dot{\sgm}\dot{\vph}_{\rm cl} \eql 
 \frac{\der V}{\der\bar{\vph}}(\vpcl)+\sum_{\vtht^*=0,\pi}
 \frac{\der\lmd_{\vtht^*}}{\der\bar{\vph}}(\vpcl)\cdot\dlt(y-R\vtht^*), 
 \label{bg_eqs}
\eea
where the dot denotes the derivative with respect to $y$, 
and $\vtht^*=0,\pi$ are the brane locations in the dimensionless 
coordinate~$\vtht$.
The first two equations come from the Einstein equation 
and the last one is the equation of motion for $\vph$. 
By integrating the above equations over infinitesimal regions 
including the boundaries~$y=y^*(=R\vtht^*)$, 
we can obtain the following jump conditions. 
\bea
 \sbk{\dot{\sgm}}_{\vtht^*} \eql 
 \left.-\frac{\kp^3}{3}\lmd_{\vtht^*}(\vpcl)\right|_{y=y^*}, \nonumber\\
 \sbk{\dot{\vph}_{\rm cl}}_{\vtht^*} \eql 
 \left.\frac{\der\lmd_{\vtht^*}}{\der\bar{\vph}}(\vpcl)\right|_{y=y^*}, 
 \label{jump_cond}
\eea
where $\sbk{\cdots}_{\vtht^*}$ is defined as 
\be
 \sbk{\alp(y)}_{\vtht^*} \equiv \alp(y^*+0)-\alp(y^*-0). \label{sbk_bd}
\ee

\section{Boundaries in the Newton gauge} \label{bd_cond_Ng}
In this appendix, we will show that the boundaries can be expressed 
by the rigid values of the coordinate~$y$ within the Newton gauge 
{\it at the linearized order}. 

To describe the boundary conditions, the Gaussian normal (GN)
coordinates are useful~\cite{montes} because the boundary 
is expressed by the rigid value of $y$. 
In this gauge, however, two coordinate patches are necessary 
for covering the whole spacetime. 
The first patch contains one boundary and the second one contains 
the other boundary whose locations are expressed 
by $y=y^+(x)$ and $y=y^-(x)$ in the Newton gauge, respectively. 
We can move from the Newton gauge to the GN gauge 
by the following transformation. 
\be
 x^\mu_{\rm GN} = x^\mu_{\rm Newton}+\xi^\mu, \label{cd_trf}
\ee
where the transformation parameters~$\xi^\mu(x,y)$ are
\bea
 \xi_y(x,y) \eql 2\int_{y^\pm}^y \dr y^\prime\; B(x,y^\prime)
 +\xi_y^{\pm}(x), \nonumber\\
 \xi_m(x,y) \eql -2\int_{y^\pm}^y \dr y^\prime\; 
 e^{-2\sgm(y^\prime)}\int_{y^\pm}^{y^\prime} \dr y''\der_m B(x,y'') 
 \nonumber\\
 &&-\brc{\der_m \xi_y^{\pm}-2\der_m y^\pm B(x,y^\pm)}
 \int_{y^\pm}^y \dr y^\prime e^{-2\sgm(y^\prime)} 
 +\xi_m^{\pm}(x).  
 \label{xi_mu}
\eea
Here, $B=\frac{1}{4}h_{yy}$ is the fluctuation of the metric~$g_{yy}$ 
around the background in the Newton gauge (See Eq.(\ref{NT_gauge}).), 
and $y^\pm(x)$ and $\xi_\mu^\pm(x)$ are functions of only $x^m$. 
From Eq.(\ref{cd_trf}) and the first equation of Eq.(\ref{xi_mu}), 
we can see that 
\be
 y^\pm(x)+\xi_y^\pm(x) = \mbox{constant}  \label{y_xi_const}
\ee 
because the boundaries are expressed by $y=\mbox{constant}$ 
in the GN coordinates. 

By using the above transformation, we can obtain 
the following boundary conditions in the Newton gauge~\cite{montes,EMS}. 
\bea
 \der_y h^{\rm TT}_{mn}+2e^{-2\sgm}\der_m\der_n\xi_y^\pm \eql 0, \nonumber\\
 \der_y B+2\dot{\sgm}B+\ddot{\sgm}\xi_y^\pm  \eql 0  \label{bd_cond_ap}
\eea
at $y=y^\pm$. 
For simplicity, we have assumed that the fluctuation~$\tl{\vph}$ vanishes 
at the boundaries. 
Taking the trace of the first condition, 
we can find that $\Box_4 \xi_y^\pm=0$. 

By solving the linearized Einstein equation, the scalar perturbation 
of the metric~$B$ is expanded into the K.K. modes. 
\be
 B(x,y) = \sum_p f_{(p)}(y)b_{(p)}(x), \label{B_mode_ex}
\ee
where each mode satisfies 
\be
 \Box_4 b_{(p)} = m_{(p)}^2 b_{(p)}+\cO(b^2). \label{KG_eq}
\ee
Here, $\Box_4\equiv\eta^{mn}\der_m\der_n$, and 
$m_{(p)}$ is the mass eigenvalue of the $p$-th K.K. mode. 
Plugging Eq.(\ref{B_mode_ex}) into the second condition 
in Eq.(\ref{bd_cond_ap}), we can obtain 
\be
 \sum_k\brkt{\der_y f_{(p)}+2\dot{\sgm}f_{(p)}}b_{(p)} = 
 -\ddot{\sgm}\xi_y^\pm. \label{sum_fb_xi} 
\ee
Operating $\Box_4$ on the both sides, 
\be
 \sum_{p^\prime}\brkt{\der_y f_{(p^\prime)}+2\dot{\sgm}f_{(p^\prime)}}
 m_{(p^\prime)}^2 b_{(p^\prime)} = 0  \label{sum_fb}
\ee
at the linearized level. 
Here, $p^\prime$ does not include the massless modes, 
\ie, $m_{(p^\prime)}^2\neq 0$. 
We have used $\Box_4 \xi_y^\pm=0$ and Eq.(\ref{KG_eq}). 

Since $b_{(p)}(x)$ are independent fields, Eq.(\ref{sum_fb}) means that 
\be
 \brkt{\der_y f_{(p^\prime)}+2\dot{\sgm}f_{(p^\prime)}}_{y=y^\pm} = 0. 
\ee
Therefore, Eq.(\ref{sum_fb_xi}) becomes 
\be
 \sum_{m_{(p)}^2=0}\brkt{\der_y f_{(p)}+2\dot{\sgm}f_{(p)}}b_{(p)} = 
 -\ddot{\sgm}\xi_y^\pm. \label{sum_fb_xi2}
\ee

On the other hand, there is no physical massless mode in $B$ 
due to the stabilization mechanism.\footnote{ 
The existence of a physical massless mode of $B$ means that 
the radius is unstabilized.}
For example, in the model of Ref.~\cite{EMS}, there are two massless mode 
solutions in $B$. 
However, one of them can be gauged away within the Newton gauge 
and the other is forbidden by the boundary condition. 
These massless mode solutions are related to $\xi_y^\pm(x)$ 
through the boundary condition~(\ref{sum_fb_xi2}). 
Hence, the absence of the massless mode means 
that we can set $\xi_y^\pm=0$ within the Newton gauge. 
Namely, $y^\pm(x)$ can be taken as constants. 
(See Eq.(\ref{y_xi_const}).) 
Note that this argument holds only {\it at the linearized level}. 
Including the higher order in the discussion, 
the boundaries cannot be expressed by $y=\mbox{constant}$ 
in the Newton gauge any more.

\section{Superfields and gauge fixing} \label{sf_gf}
Here, we will collect explicit forms of the $\cN=1$ superfields 
in terms of the superconformal notation of Ref.\cite{KO1,KO2}. 

\subsection{5D superfields} \label{def_sf}
A vector multiplet~$\cV^I$ consists of 
\be
 M^I,\;\;\; W^I_\mu,\;\;\; \Omg^{Ii},\;\;\; Y^{I(r)}, 
\ee
which are a gauge scalar, a gauge field, a gaugino 
and an auxiliary field, respectively. 
The indices~$i=1,2$ and $r=1,2,3$ are doublet and triplet indices 
for $\SUu$. 
From this multiplet, we can define 
the following vector and chiral superfields. 
\bea
 V^I \defa \tht\sgm^{\udl{m}}\btht W^I_m+i\tht^2\btht\bar{\lmd}^I
 -i\btht^2\tht\lmd^I+\frac{1}{2}\tht^2\btht^2D^I, \nonumber\\
 \Phi^I_S \defa \vph_S^I-\tht\chi_S^I-\tht^2\cF_S^I, 
\eea
where
\bea
 \lmd^I \defa 2e^{\frac{3}{2}F}\Omg^{I1}_{\rm R}, \nonumber\\
 D^I \defa -e^{2F}\brc{G^{-1}\der_yM^I-2Y^{I(3)}
 +G^{-1}\dot{F}M^I}, \nonumber\\
 \vph_S^I \defa \frac{1}{2}\brkt{W_y^I+iG M^I}, \nonumber\\
 \chi_S^I \defa -2e^{\frac{F}{2}}G\Omg^{I2}_{\rm R}, \nonumber\\
 \cF_S^I \defa -ie^F G\brkt{Y^{I(1)}+iY^{I(2)}}. 
 \label{sf_comp1}
\eea

The hypermultiplets consist of complex scalars~$\cA_i^\alp$, 
spinors~$\zeta^\alp$ and auxiliary fields~$\cF_i^\alp$. 
They carry a $\Usp$ index~$\alp$ ($\alp=1,2,\cdots,2\nH+2$) 
on which the gauge group can act. 
These are split into $\nH+1$ hypermultiplets as 
\be
 \cH^{\hat{\alp}} = \brkt{\cA_i^{2\hat{\alp}+1}, 
\cA_i^{2\hat{\alp}+2},\zeta^{2\hat{\alp}+1},\zeta^{2\hat{\alp}+2}, 
\cF_i^{2\hat{\alp}+1},\cF_i^{2\hat{\alp}+2}}. 
\ee 
From these multiplets, we can define the following chiral superfields. 
\be
 \Phi^\alp \equiv \vph^\alp-\tht\chi^\alp-\tht^2\cF^\alp, 
\ee
where
\bea
 \vph^\alp \defa \cA^\alp_2, \nonumber\\
 \chi^\alp \defa -2ie^{\frac{F}{2}}\ztR^\alp,  \nonumber\\
 \cF^\alp \defa e^FG^{-1}\brc{\der_y\cA_1^\alp
  +i\brkt{G+\frac{iW_y^0}{M^0}}\tl{\cF}_1^\alp 
  -\brkt{W_y^I-iG M^I}(gt_I)^\alp_{\;\;\bt}\cA_1^\bt
  +\frac{3}{2}\dot{F}\cA_1^\alp}. \nonumber\\
 \label{sf_comp2}
\eea
In the last expression, $\tl{\cF}_1^\alp$ are defined as
\be
 \tl{\cF}_1^\alp \equiv \cF_1^\alp-M^0(gt_0)^\alp_{\;\;\bt}\cA_1^\bt. 
\ee

Since we treat the radion supermultiplet separately, 
we have not included the dependence on the $\SUu$ gauge field~$V_y^{(r)}$, 
which corresponds to an auxiliary field of the radion multiplet, 
in the above definitions of the superfields.

\subsection{4D superfields} \label{4Dsf}
We can construct superfields also from 4D superconformal multiplets. 
From a vector multiplet~$(B_m^A,\lmd^A,\bar{\lmd}^A,D^A)$, 
we can obtain the following vector superfield. 
\be
 V^A \equiv \tht\sgm^{\udl{m}}\bar{\tht}B_m^A
 +ie^{\frac{3}{2}F(y^*)}\tht^2\bar{\tht}\bar{\lmd}^A
 -ie^{\frac{3}{2}F(y^*)}\bar{\tht}^2\tht\lmd^A
 +\frac{1}{2}e^{2F(y^*)}\tht^2\bar{\tht}^2 D^A, 
\ee
where $y^*=0,\pi R$ are the brane locations. 
The corresponding superfield strength is defined as 
\be
 \cW_\alp^A \equiv -\frac{1}{4}\bar{D}^2 D_\alp V^A. 
\ee

From a chiral multiplet~$(s^a,\chi_s^a,\cF_s^a)$, we can construct 
the following chiral superfield. 
\be
 S^a \equiv s^a-e^{\frac{F}{2}(y^*)}\tht\chi_s^a-e^{F(y^*)}\tht^2\cF_s^a. 
\ee

The warp factors in the above definitions come from 
the induced metric on the boundaries.

\subsection{Superconformal gauge fixing} \label{sc_gf}
The gauge fixing conditions for the extraneous superconformal symmetries, \ie, 
the dilatation~$\bdm{D}$, $\SUu$, the conformal 
supersymmetry~$\bdm{S}$ are as follows.\footnote{
The special conformal transformation~$\bdm{K}$ 
is already fixed in our superspace formalism \cite{AS}. 
}

The $\bdm{D}$-gauge is fixed by 
\bea
 \cN \defa C_{IJK}M^IM^JM^K = M_5^3, \nonumber\\
 \cA_i^\alp\dmx\cA^i_\bt \eql 2\brc{-\sum_{a=1}^2\abs{\cA_2^a}^2
 +\sum_{\udl{\alp}=3}^{2\nH+2}\abs{\cA_2^{\udl{\alp}}}^2} = -2M_5^3, 
\eea
where $\cN$ is called the norm function, and $\SUu$ is fixed 
by the condition 
\be
 \cA_i^\alp \propto \dlt_i^\alp, \;\;\;\;\;
 (\alp=1,2). 
\ee

The $\bdm{S}$-gauge is fixed by 
\bea
 \cN_I\Omg^{Ii} \eql 0, \nonumber\\
 \cA_i^\alp\dmx\zeta_\bt \eql 0, 
\eea
where $\cN_I\equiv\der\cN/\der M^I$.

\end{document}